\documentclass[a4paper,11pt]{article}
\usepackage[dvipdfmx]{graphicx}
\usepackage{amsfonts}
\usepackage{physics,amsmath}
\usepackage{amssymb}
\usepackage{mathtools}
\usepackage{colortbl}
\usepackage[table]{xcolor}
\usepackage{tikz}
\usetikzlibrary{shapes.geometric}
\usetikzlibrary {shapes.misc}
\usepackage{cite}
\usepackage{here}
\usepackage{hyperref}
\usepackage{mathrsfs}
\usepackage{algpseudocode}
\usepackage{ulem}
\usepackage{bm}
\usepackage{makecell}
\usepackage{comment}
\numberwithin{equation}{section}
\usepackage[left=2cm,right=2cm,top=3.5cm,bottom=3.5cm]{geometry}

\allowdisplaybreaks[1]
\hypersetup{
	colorlinks=true,
	linkcolor=ceruleanblue,
	filecolor=ceruleanblue,      
	urlcolor=ceruleanblue,
	citecolor=ceruleanblue,
}
\definecolor{ceruleanblue}{rgb}{0.0, 0.2, 0.6}
\newcommand{\de}{{\rm d}}

\let\originalleft\left
\let\originalright\right
\renewcommand{\left}{\mathopen{}\mathclose\bgroup\originalleft}
\renewcommand{\right}{\aftergroup\egroup\originalright}

\date{\today}

\begin{document}
\begin{flushright} {\footnotesize YITP-25-79, IPMU25-0028, RIKEN-iTHEMS-Report-25} \end{flushright}

\begin{center}
\LARGE{\bf Parametrized Tidal Dissipation Numbers of Non-rotating Black Holes}
\\[1cm] 

\large{Hajime Kobayashi$^{\,\rm a}$, Shinji Mukohyama$^{\,\rm a, \rm b, \rm c}$, Naritaka Oshita$^{\,\rm a, \rm d, \rm e}$, Kazufumi Takahashi$^{\,\rm f, \rm a}$,\\
and Vicharit Yingcharoenrat$^{\,\rm g, \rm c}$}
\\[0.5cm]

\small{\textit{$^{\rm a}$
Center for Gravitational Physics and Quantum Information, Yukawa Institute for Theoretical Physics, 
\\ Kyoto University, 606-8502, Kyoto, Japan}}
\vspace{.2cm}

\small{\textit{$^{\rm b}$
Research Center for the Early Universe (RESCEU), Graduate School of Science, The University of Tokyo,
\\ Hongo 7-3-1, Bunkyo-ku, Tokyo 113-0033, Japan}}
\vspace{.2cm}

\small{
\textit{$^{\rm c}$
Kavli Institute for the Physics and Mathematics of the Universe (WPI), The University of Tokyo Institutes for Advanced Study (UTIAS), The University of Tokyo, Kashiwa, Chiba 277-8583, Japan}}
\vspace{.2cm}

\small{
\textit{$^{\rm d}$
The Hakubi Center for Advanced Research, Kyoto University, Yoshida Ushinomiyacho, Sakyo-ku, Kyoto 606-8501, Japan}}
\vspace{.2cm}

\small{
\textit{$^{\rm e}$
RIKEN iTHEMS, Wako, Saitama, 351-0198, Japan}}
\vspace{.2cm}

\small{
\textit{$^{\rm f}$
Department of Physics, College of Humanities and Sciences, Nihon University, Tokyo 156-8550, Japan}}
\vspace{.2cm}

\small{
\textit{$^{\rm g}$
High Energy Physics Research Unit, Department of Physics, Faculty of Science, Chulalongkorn University, Pathumwan, Bangkok 10330, Thailand}}
\vspace{.2cm}
\end{center}

\vspace{0.3cm} 

\begin{abstract}\normalsize
A set of tidal dissipation numbers (TDNs) quantifies the absorption of the tidal force exerted by a companion during an inspiralling phase of a binary compact object.
This tidal dissipation generally affects the gravitational waveform, and measuring the TDNs of a black hole (BH) allows us to test the nature of gravity in the strong-field regime.
In this paper, we develop a parametrized formalism for calculating the TDNs of static and spherically symmetric BH backgrounds using the Mano-Suzuki-Takasugi method, which connects the underlying perturbative equations with observable quantities in gravitational-wave observations in a theory-agnostic manner.
Our formalism applies to any system where the master equation has the form of the Regge-Wheeler/Zerilli equation with a small correction to the effective potential.
As an application of our formalism, we consider three examples: the effective field theory of BH perturbations with timelike scalar profile, the Einstein-Maxwell system, and a higher-curvature extension of general relativity.
We also discuss the absence of logarithmic running for the TDNs.
\end{abstract}

\vspace{0.3cm} 

\vspace{2cm}

\newpage
{
\hypersetup{linkcolor=black}
\tableofcontents
}

\flushbottom

\section{Introduction and summary}\label{sec:Introduction}
Tidal fields serve as powerful and insightful probes for understanding the internal structure of compact objects, the fundamental nature of gravity, and the environmental fields (i.e., ``hair'') around the objects. 
Tidal interactions between binary constituents affect the orbital dynamics and, in principle, leave detectable imprints on gravitational wave (GW) signals. 
Bridging the gap between tidal effects predicted by fundamental theories and observations of GWs requires accurate and theory-agnostic modelings.

In the late stages of inspiral, compact objects such as black holes (BHs) can be deformed by external tidal fields. 
These deformations encode information about the properties of compact objects and the underlying theory of gravity, and are captured by their {\it tidal response functions}. 
The leading conservative tidal effects are quantified by the so-called tidal Love numbers (TLNs)~\cite{Hinderer:2007mb,Damour:2009vw,Binnington:2009bb}, which characterize how easily an object deforms in response to an external tidal field.
TLNs give rise to phase corrections to the emitted GWs, starting at 5 post-Newtonian (5PN) order ~\cite{Flanagan:2007ix,Damour:2012yf,Favata:2013rwa}. These effects were constrained by using data from the GW event~GW170817~\cite{LIGOScientific:2017vwq,LIGOScientific:2018hze,LIGOScientific:2018cki}.
In four-dimensional general relativity (GR), it was shown that the vacuum solutions, e.g., Kerr BHs, have vanishing TLNs for generic multipole modes with $\ell \geq 2$~\cite{Damour:2009vw,Binnington:2009bb,Kol:2011vg,LeTiec:2020spy,Hui:2020xxx,Chia:2020yla,LeTiec:2020bos,Charalambous:2021mea}.
However, BHs in theories beyond four-dimensional GR generally can have non-vanishing TLNs~\cite{Cardoso:2017cfl,Cardoso:2018ptl,Chakravarti:2018vlt,Cardoso:2019vof,Brown:2022kbw,DeLuca:2022tkm,Katagiri:2023umb,Barura:2024uog,Barbosa:2025uau,Cano:2025zyk}.
In addition, horizonless compact objects that may arise due to quantum gravity corrections~(see, e.g.,~\cite{Uchikata:2016qku,Cardoso:2017cfl,Addazi:2018uhd,Maselli:2018fay,Cardoso:2019rvt,Cardoso:2019nis,Chakraborty:2023zed} and references therein) or BHs surrounded by matter fields in general possess non-zero TLNs~\cite{Cardoso:2019upw,DeLuca:2021ite,DeLuca:2022xlz,Katagiri:2023yzm,Katagiri:2023umb}.
In fact, TLNs can also exhibit a logarithmic scale dependence, indicating ``running'' TLNs (see, e.g.,~\cite{Cardoso:2019upw,Hui:2020xxx,DeLuca:2022tkm,DeLuca:2022xlz,Katagiri:2023umb,Barbosa:2025uau} and references therein).
The detection of non-vanishing TLNs of BHs would provide compelling evidence for deviations from GR in vacuum.
These deviations can be analyzed within a parametrized formalism~\cite{Katagiri:2023umb}, as long as they are described by a decoupled master equation with a single degree of freedom.\footnote{See \cite{Cardoso:2019mqo,McManus:2019ulj} for the parametrized BH quasinormal ringdown formalism, which applies to Regge-Wheeler/Zerilli equations with small corrections.
See also \cite{Cano:2024jkd} for an extension of the formalism to modified Teukolsky equations~\cite{Li:2022pcy,Li:2023ulk}.}

In addition to the conservative tidal effects, there are dissipative tidal responses that lead to energy loss and affect the phase of gravitational waveforms.
These effects enter at 4PN order for non-rotating objects and at 2.5PN order for spinning objects~\cite{Poisson:1994yf,Tagoshi:1997jy, Alvi:2001mx,Poisson:2004cw}.
These dissipative effects are quantified by the so-called {\it tidal dissipation numbers} (TDNs)~\cite{Ivanov:2022hlo,Katagiri:2023yzm,Katagiri:2024fpn}.
In the case of BHs, the dissipation process is attributed to the absorption of gravitational perturbations by the event horizon.\footnote{In the case of neutron stars, TDNs account for energy loss due to viscous dissipation (bulk/shear viscosity) of the internal fluids (see \cite{Ghosh:2023vrx,Ripley:2023qxo,HegadeKR:2024agt} and references therein).}
Thus, measuring TDNs allows us to observationally distinguish BHs from horizonless compact objects that may arise from Planckian corrections~\cite{Maselli:2017cmm,Addazi:2018uhd,Cardoso:2019rvt,Datta:2019epe,Datta:2020rvo,Maggio:2021uge,Chakraborty:2023zed}.
TDNs were calculated both within four-dimensional GR~\cite{Chia:2020yla,Charalambous:2021mea,Ivanov:2022hlo,Ivanov:2022qqt,Saketh:2022xjb,Saketh:2023bul,Chakraborty:2023zed,Perry:2023wmm,Katagiri:2023yzm,Bautista:2023sdf,Katagiri:2024wbg} and extensions beyond GR~\cite{Katagiri:2024fpn}.
It is important to note that in extreme mass ratio inspirals, even small deviations in TDNs can accumulate over thousands of orbital cycles, leaving detectable imprints on the gravitational waveform (see e.g., \cite{Datta:2019epe,Datta:2019euh,Datta:2020rvo,Maggio:2021uge,Datta:2024vll}).
In the case of comparable-mass binaries, the importance of tidal dissipation effects in distinguishing between BHs and neutron stars has been pointed out~\cite{Datta:2020gem,Mukherjee:2025wxa}, and the measurability in the current and next-generation observations is discussed in \cite{Mukherjee:2022wws,Chia:2024bwc,Shterenberg:2024tmo}.

Unlike the case of TLNs, a parametrized BH TDN formalism has not yet been developed in the literature. 
In this paper, we establish for the first time such a parametrized framework for BH TDNs, revealing their general properties in a model-agnostic way.
Our approach is based on introducing perturbative modifications to the master equations---namely, the Regge-Wheeler/Zerilli equations~\cite{Regge:1957td,Zerilli:1970wzz}---which describe the dynamics of metric perturbations around a static and spherically symmetric BH.
In order to analytically solve the perturbation equations, we use the Mano-Suzuki-Takasugi (MST) method~\cite{Mano:1996vt,Mano:1996mf,Mano:1996gn}, and examine how small deviations in the effective potential change the resulting physical observables. 
By performing the asymptotic matching between the MST horizon-ingoing solution and the far-zone perturbations obtained via the worldline effective field theory (EFT) approach (see, e.g., \cite{Chakrabarti:2013lua,Hui:2020xxx,Charalambous:2021mea}), we unambiguously extract the static TLNs and TDNs. 
In this framework, we use the so-called renormalized angular momentum to clearly distinguish between the tidal source and the induced multipole moment. 
This procedure ensures that the static TLNs and TDNs are uniquely determined, and remain unaffected by ${\cal O}(1/r)$ corrections arising from the redefinition of the master variable~\cite{Cardoso:2019mqo,Barura:2024uog}.

\paragraph{Outline and summary}
Here, we summarize the outline and results of our paper.
Our main results are given by Eqs.~(\ref{eq:TLN_TDN_para}) and (\ref{eq:TLN_TDN_even_para}) together with those shown in Tables~\ref{tae_e(1)}, \ref{table(1)_num_odd}, and \ref{table(1)_even}.
These are obtained via the following step-by-step workflow:
\begin{enumerate}
    \item In Section~\ref{sec:review-tidal-response}, we review the definition of the tidal response function [Eq.~(\ref{def-bare-tidal-MST})] from the asymptotic behaviors of the near-zone solution using the MST method. 
    Based on this definition, we can unambiguously extract the TDNs of a Schwarzschild BH in the odd-parity sector as in Eq.~(\ref{eq:TDN_Sch_GR}).
    \item In Section~\ref{sec:parametrized-formalism}, we first define our parametrized TDN formalism based on the deformed Zerilli/Regge-Wheeler equations~\eqref{deformed-RWZ}.
    Here, deviations from the vacuum GR, appearing in the modifications of the effective potential~\eqref{deformed-potential}, are controlled by small coefficients~$\alpha_j^\pm$ in the $r_g/r$-expansion, where $r_g$ is the horizon radius and the superscript~$\pm$ corresponds to the parity even/odd.
    \item We then solve Eq.~\eqref{deformed-RWZ} up to first order in $\alpha_j^\pm$ and in the frequency~$\omega$, using the Green's function method.
    By using the vacuum-GR Green's function [Eq.~(\ref{eq:Green_fn})], the first-order solutions in the odd-parity sector are formally given by Eq.~\eqref{eq:GRW-inhomogeneous-sol}.
    \item Using the asymptotic behaviors of the first-order solutions [Eq.~(\ref{asympt})], we define the modification of the response function as in Eq.~(\ref{bare-psi}). This can be written in terms of the bases~$e^{-,\,j}_{(0)}$ and $e^{-,\,j}_{(1)}$ defined in Eq.~(\ref{odd-basis}). 
    Analytic expressions and numerical values of $e^{-,\,j}_{(1)}$ for $\ell = 2, 3, 4$ are presented in Tables~\ref{tae_e(1)} (with $j \geq 3$) and \ref{table(1)_num_odd} (with $3 \leq  j \leq 9 $), respectively.
    \item Finally, from Eq.~(\ref{deviated-bare-response-odd}), we read off the static TLNs and the TDNs for the odd-parity sector as in Eq.~(\ref{eq:TLN_TDN_para}).
    A similar methodology can be straightforwardly applied to the even-parity sector, resulting in Eq.~(\ref{eq:TLN_TDN_even_para}). 
    Numerical values of $e^{+,\,j}_{(1)}$ for $\ell = 2, 3, 4$ with $3 \leq  j \leq 9 $ are shown in Table~\ref{table(1)_even}.
\end{enumerate}
After obtaining our main results, in Section~\ref{sec:applications}, we apply our formalism to compute odd-parity TDNs in three examples: the EFT of BH perturbations with timelike scalar profile~\cite{Mukohyama:2022enj,Mukohyama:2022skk}, the Einstein-Maxwell system, and the higher curvature-extension of GR~\cite{Endlich:2017tqa,Cardoso:2018ptl}.
Finally, we draw our conclusions in Section~\ref{sec:conclusions}.
Throughout this paper, we use the unit~$c=G=1$.

\section{Formulation of the tidal response via MST method}\label{sec:review-tidal-response}
In this section, we explain the relativistic formulation of TDNs in four-dimensional GR, using the MST method. 
In order to compute the TDNs, one needs to properly define the tidal response function.
In fact, it was pointed out in the literature~\cite{Damour:2009vw,Binnington:2009bb,Kol:2011vg,Damour:2009va,Gralla:2017djj,Steinhoff:2021dsn,Katagiri:2023umb} that there are ambiguities in the definition of the tidal response function in the relativistic formulation, as opposed to the Newtonian case~\cite{Binnington:2009bb,Poisson2014gravity,Charalambous:2021mea}.
Such ambiguities arise when the growing mode of metric perturbations acquires relativistic corrections or overlaps with the decaying mode~\cite{Gralla:2017djj,Steinhoff:2021dsn,Charalambous:2021mea,Barura:2024uog}.
One way to resolve these ambiguities is to analytically continue the multipole index~$\ell$ to non-integer values~\cite{Kol:2009mj,Chakrabarti:2013lua,Steinhoff:2014kwa,Hui:2020xxx,Chia:2020yla,LeTiec:2020bos,Creci:2021rkz}.
In addition, as discussed in \cite{Chia:2020yla,Steinhoff:2021dsn}, the ambiguity issue can be resolved when properly imposing the ingoing and outgoing boundary conditions at the horizon and at infinity.

Here, we employ the MST method in the frequency domain to unambiguously define the tidal response function. 
As we will show below, in this approach, the solutions of the Regge-Wheeler/Zerilli equation are expressed in the form of a series of hypergeometric functions in the near zone and a series of the Coulomb wavefunctions in the far zone.
These series are characterized by the renormalized angular momentum~$\nu$, which is a complex quantity that allows us to distinguish between the source and response terms~\cite{Chakrabarti:2013lua,Charalambous:2021mea,Perry:2024vwz}.
Note that $\nu$ can be interpreted as an analytic continuation of $\ell$ to a complex number, as $\nu \to \ell$ in the limit of $\omega \to 0$.

Throughout this section, we assume that the spacetime metric is given by that of the Schwarzschild BH,
\begin{align}\label{eq:bg_Sch}
    \bar{g}_{\mu\nu}\de x^\mu \de x^\nu=-f(r) \de t^2+\frac{\de r^2}{f(r)}+r^2(\de\theta^2+\sin^2\theta\,\de\varphi^2)\;,
\end{align}
where $f(r)\equiv 1-r_g/r$ and $r_g$ denotes the horizon radius.

\subsection{Separation of scales}\label{ssec:scale}
Before analyzing BH perturbations based on the MST method, we briefly discuss a scale for matching to extract the tidal source and induced response.
The description of the tidal response in a relativistic regime relies on an expansion between the vacuum neighborhood of the body and a nearby Universe, except for the body's vicinity.
Let us consider a two-body system composed of a (non-rotating) BH and its companion with the orbital separation length~$r_{\rm orb}$, the typical orbital frequency~$\omega$, and the relative velocity~$v_{\rm orb} \equiv \omega r_{\rm orb}$.
We assume (\textit{i})~large separation ($r_{\rm orb}\gg r_g$) and
(\textit{ii})~the adiabatic evolution of tidal environments ($v_{\rm orb} \ll 1$).
If the location we are interested in lies outside the horizon but remains much closer to the BH than to its companion, the tidal field can be treated as a perturbation to the BH.
This approximation is valid in the regime~$r_g<r\ll r_{\rm orb}$.
The information on the PN potential as a two-body problem is contained in the region of $r_g\ll r\sim r_{\rm orb}\ll \omega^{-1}$.
By performing a matching in the overlap region~$r_g\ll r\ll r_{\rm orb}\ll \omega^{-1}$, one can extract the near-horizon information from the far-zone observables, e.g., the orbital evolution of the companion or GW signals.\footnote{Note that the scale of GW emission is $\lambda_{\rm GW}\sim\omega^{-1}\gg r_{\rm orb}$.}
In what follows, we will analyze the tidal response using the horizon-ingoing solution obtained from the MST method.

\subsection{Tidal response from MST method}
Here, we focus on the dynamics of linear perturbations on the Schwarzschild background~\eqref{eq:bg_Sch} in GR.
As usual, with the use of spherical harmonics decomposition, the linearized Einstein equations give rise to equations of the radial components for each multipole.
When imposing the Regge-Wheeler gauge, the radial perturbations in the odd- and even-parity sectors are described in the frequency domain by the standard Regge-Wheeler and Zerilli equations~\cite{Regge:1957td,Zerilli:1970wzz}:
\begin{align}
    \left[\left(f(r)\frac{\de}{\de r}\right)^2+\omega^2-f(r)V_\ell^{\pm}(r)\right]\psi_\ell^\pm(r)=0\;, \label{GR-RW}
\end{align}
where $\psi_\ell^\pm(r)$ are the master variables and the effective potentials~$V_\ell^{\pm}(r)$ are given by
\begin{equation}
\begin{split}
    V_\ell^+(r) &\equiv \frac{9\lambda r_g^2 r+3\lambda^2 r_g r^2+\lambda^2(\lambda+2)r^3+9r_g^3}{r^3(\lambda r+3r_g)^2}\;,\\
    V_\ell^-(r) &\equiv \frac{\ell(\ell+1)}{r^2}-\frac{3r_g}{r^3}\;,
\end{split}
\end{equation}
with $\lambda\equiv \ell(\ell+1)-2$.
We use the superscript~``$+$'' to denote quantities defined in the even-parity sector, whereas ``$-$'' stands for those in the odd-parity sector.
Note that the radial equations do not depend on the azimuthal magnetic quantum number~$m$ due to the spherical symmetry of the background.

The homogeneous equation~\eqref{GR-RW} has two independent solutions:  $\psi_\ell^{\pm {\rm in}}(r)$, which is purely ingoing at the horizon~$r = r_g$, and  $\psi_\ell^{\pm {\rm up}}(r)$, which is purely outgoing at infinity.
For the odd-parity sector, these homogeneous solutions can be obtained using the MST formalism~\cite{Mano:1996gn,Mano:1996mf,Mano:1996vt} (see also \cite{Sasaki:2003xr} for a review) in the low-frequency regime ($r_g\omega\ll 1$).
Then, after obtaining the asymptotic behavior of the horizon-ingoing solution in the buffer zone ($r_g\ll r\ll r_{\rm orb}\ll \omega ^{-1}$), we can extract the low-frequency expansion of the tidal response function and the TDNs.
We will review this analysis in Sections~\ref{sssec:near-zone-MST}--\ref{ssec:tidal-MST}.
For the even-parity sector, it was shown in Refs.~\cite{Forseth:2015oua,Munna:2020iju} that the solution can be derived by use of the Chandrasekhar transformation~\cite{Chandrasekhar:1975nkd,Chandrasekhar:1975zz,Chandrasekhar:1985kt}.
We will discuss this point in Section~\ref{sec:even_transformation}.

\subsubsection{Near-zone solution in the odd-parity sector}\label{sssec:near-zone-MST}
Here, we apply the MST method to obtain the solutions of the Regge-Wheeler equation. 
In particular, we will discuss the solutions in the near-zone limit ($r_g < r \ll \omega^{-1}$).
For later convenience, we define the dimensionless radial coordinate~$x\equiv r/r_g$ and the dimensionless frequency~$\varepsilon\equiv r_g\omega$.

For the horizon-ingoing solution to the Regge-Wheeler equation~\eqref{GR-RW}, we consider the following ansatz:
\begin{align}
    \psi_\nu^{- {\rm in}}(x)&=e^{-i\varepsilon x}\left(1-\frac{1}{x}\right)^{-i\varepsilon}\sum_{n=-\infty}^\infty a_n^\nu\,p_{n+\nu}(x)\;, \label{MST-hor-in}
\end{align}
where $p_{n+\nu}(x)$ is the hypergeometric function specified by an auxiliary parameter~$\nu$, called the renormalized angular momentum, as
\begin{equation}
\begin{split}
    p_{n+\nu}(x)&=\frac{\Gamma(g)\Gamma(h)}{\Gamma(k)} {}_2F_1\left(g, h; k;1-x\right)\;,
\end{split}
\end{equation}
with $g \equiv n+\nu-1-i\varepsilon,\ h \equiv -n-\nu-2-i\varepsilon$, and $k \equiv 1-2i\varepsilon$.
Here, ${}_2F_1$ denotes the ordinary hypergeometric function and $\Gamma$ stands for the Gamma function. 
Note that the factor in front of the series in Eq.~(\ref{MST-hor-in}) has been chosen so that the function is regular at $r = 0$ and $r = r_g$.
Substituting the ansatz~(\ref{MST-hor-in}) into the Regge-Wheeler equation~\eqref{GR-RW}, we obtain the following three-term recurrence relation for the coefficients~$a_n^\nu$:
\begin{align}
    \alpha_n^\nu a_{n+1}^\nu+\beta_n^\nu a_n^\nu+\gamma_n^\nu a_{n-1}^\nu=0\; , \label{3-term-reccurence}
\end{align}
where the coefficients~$\alpha_n^\nu$, $\beta_n^\nu$, and $\gamma_n^\nu$ are given by
\begin{equation}\label{alpha-beta-gamma}
\begin{split}
    \alpha_n^\nu&=-i\varepsilon \frac{g(g+2)(-h-3)}{(n+\nu+1)(g-h+2)}\;,\\
    \beta_n^\nu&=(n+\nu)(n+\nu+1)-\ell(\ell+1)+2\varepsilon^2+\frac{\varepsilon^2(\varepsilon^2+4)}{(n+\nu)(n+\nu+1)}\;,\\
    \gamma_n^\nu&=i\varepsilon \frac{h (h+2)(g+3)}{(n+\nu)(g-h-2)}\;.
\end{split}
\end{equation}
By introducing $R_n(\nu)\equiv a_n^\nu/a_{n-1}^\nu$ and $L_n(\nu)\equiv a_n^\nu/a_{n+1}^\nu$, Eq.~\eqref{3-term-reccurence} can be written in the form of infinite continued fraction:
\begin{equation}
    R_n(\nu)= - \frac{\gamma_n^\nu}{\displaystyle \beta_n^\nu-\frac{\alpha_n^\nu\gamma_{n+1}^\nu}{\beta_{n+1}^\nu-\cdots}}\;,\quad
    L_n(\nu)=-\frac{\alpha_n^\nu}{\displaystyle \beta_n^\nu-\frac{\gamma_n^\nu\alpha_{n-1}^\nu}{\beta_{n-1}^\nu-\cdots}}\;,
    \label{reccurence-for-RL} 
\end{equation}
for a fixed $n$.
For each of these two continued fractions, one can obtain $\{a_n^\nu\}$ so that the infinite continued fraction converges.
Note that the two sets of coefficients~$\{a_n^\nu\}$ [i.e., the one that makes $R_n(\nu)$ converge and the one that makes $L_n(\nu)$ converge] do not coincide in general.
One can make both infinite continued fractions converge by choosing $\nu$ so that the following condition is satisfied:
\begin{align}
    R_n(\nu)L_{n-1}(\nu)=1\;.\label{determine-nu}
\end{align}
There is a solution to this equation such that $\nu\to\ell$ as $\varepsilon\to 0$, which can be expressed as $\nu=\ell+\sum_{n=1}^\infty \ell_{2n}\varepsilon^{2n}$.
For instance, the coefficient of $\varepsilon^2$ can be written explicitly as~\cite{Mano:1996mf,Zhang:2013ksa,Casals:2015nja}
\begin{equation}
    \ell_2(\ell)=-\frac{15 \ell^4+30 \ell^3+28 \ell^2+13 \ell+24}{2 \ell (\ell+1) (2 \ell-1) (2 \ell+1) (2 \ell+3)}\;. \label{nu-expansion}
\end{equation}
It should be noted that the renormalized angular momentum gives contributions of ${\cal O}(\varepsilon^2)$ or higher. 
Thus, when restricting to the TDNs at ${\cal O}(\varepsilon)$, corrections of $\ell_2,\ell_4,\cdots$ do not show up.

Then, using the connection formula of the hypergeometric function, the horizon-ingoing solution~(\ref{MST-hor-in}) can be written as
\begin{align}
    \psi_\nu^{- {\rm in}}(x)=\psi_\nu^{(0)}(x)+\psi_{-\nu-1}^{(0)}(x)\;,\label{MST-hor-in-large-x}
\end{align}
where $\psi_\nu^{(0)}(x)$ is given by
\begin{equation}
    \psi_\nu^{(0)}(x) =e^{-i\varepsilon x}\left(1-\frac{1}{x}\right)^{-i\varepsilon}x^{\nu+1}\sum_{n=-\infty}^\infty a_n^\nu x^n\frac{\Gamma(h)\Gamma(g-h)}{\Gamma(g+4)} {}_2F_1\left(h,h+4;-g+h+1;\frac{1}{x}\right)\;.
\label{growing}
\end{equation}
Note that there is a symmetry under $(n,\nu)\mapsto(-n,-\nu-1)$, and therefore $a_{-n}^{-\nu-1}$ satisfies the same recurrence relation as $a_n^\nu$.
By setting the initial condition~$a_0^\nu=a_0^{-\nu-1}=1$, we have 
\begin{align}\label{eq:a_n_-nu}
    a_n^\nu=a_{-n}^{-\nu-1}\;.
\end{align}
Note also that, if $\nu$ is a solution of Eq.~\eqref{determine-nu}, then $-\nu-1$ is also a solution.
We see from Eq.~(\ref{eq:a_n_-nu}) that the decaying mode~$\psi_{-\nu-1}^{(0)}(x)$, defined in Eq.~(\ref{MST-hor-in-large-x}), becomes
\begin{equation}
    \begin{split}
    \psi_{-\nu-1}^{(0)}(x)
    &=e^{-i\varepsilon x}\left(1-\frac{1}{x}\right)^{-i\varepsilon}x^{-\nu}\sum_{n=-\infty}^\infty a_n^{\nu} x^{-n}\frac{\Gamma(g)\Gamma(-g+h)}{\Gamma(h+4)} {}_2F_1\left(g,g+4;g-h+1;\frac{1}{x}\right)\;.
    \end{split}\label{decaying}
\end{equation}
The low-frequency behaviors of $a_n^\nu$ for some $n$'s are summarized in Table~\ref{tab:MST-coeff} in Appendix~\ref{app:MST}.
So far, we have separated the horizon-ingoing solution into the growing mode~\eqref{growing} and the decaying mode~\eqref{decaying}.
Below, we will discuss the far-zone solution to be matched with the near-zone solution obtained above. In particular, we will define the in- and out-mode solutions at infinity.

\subsubsection{Far-zone (radiative) solution}\label{sssec:far-zone-MST}
Let us now analyze the solution in the far-zone region ($r \gtrsim \omega^{-1}$).
In this case, the solution behaves like a wave and it can be written in the form of a series of the Coulomb wave functions~$\psi^{\rm C}_\nu(x)$.
To do so, it is useful to point out the relation between $\psi_\nu^{(0)}(x)$ and $\psi^{\rm C}_\nu(x)$~\cite{Mano:1996gn,Mano:1996mf}, that is
\begin{align}
    \psi_\nu^{(0)}(x) =K_\nu\psi^{\rm C}_\nu(x)\;, \label{eq:psi_nu_psi_C}
\end{align}
where $K_\nu$ is a constant factor determined by matching their expansion coefficients.\footnote{The validity of this method is guaranteed by the fact that when expanding these solutions in the Laurent series of $\varepsilon x$, they both give the series with the same exponent at $x \to \infty$.}
The analytic solution for $\psi^{\rm C}_\nu(x)$ is given by 
\begin{align}
       \psi^{\rm C}_\nu(x) &= 2^\nu e^{-i\varepsilon x}\left(1-\frac{1}{x}\right)^{-i \varepsilon} (\varepsilon x)^{\nu+1} \nonumber\\
    &\quad \times \sum_{n=-\infty}^\infty a_n^\nu\frac{\Gamma(g)\Gamma(g+2)}{\Gamma(g-h+1)\Gamma(-h+1)} (2i \varepsilon x)^n\,{}_1F_1(-h-1, g - h+1; 2i\varepsilon x)\;. \label{eq:psi_C_sol}
\end{align}
The function~${}_1F_1$ denotes the confluent hypergeometric function of the first kind, which is regular at $\varepsilon x=0$.
Therefore, using Eq.~(\ref{eq:psi_nu_psi_C}), the horizon-ingoing solution~$\psi_\nu^{- {\rm in}}(x)$ is
\begin{align}
    \psi_\nu^{- {\rm in}}(x)&=K_\nu\psi^{\rm C}_{\nu}(x)+K_{-\nu-1}\psi^{\rm C}_{-\nu-1}(x)\;.\label{hor-in-coulomb-wave}
\end{align}
We note that the ratio between $K_\nu$ and $K_{-\nu-1}$ reflects the boundary condition of $\psi_\nu^{- {\rm in}}(x)$ at the horizon.
At the leading order in $\varepsilon$, the ratio~$K_{-\nu-1}/K_\nu$ is given by~\cite{Casals:2015nja}
\begin{equation}
\begin{split}
    \frac{K_{-\nu-1}}{K_\nu}&=\frac{i\sqrt{\pi}}{2}(-\varepsilon^2)^\ell\frac{\Gamma(\ell+1)\Gamma^2(\ell-1)\Gamma^2(\ell+3)}{\ell_2^2\Gamma(2\ell+1)\Gamma(\ell+\frac{1}{2})\Gamma^2(2\ell+2)}+{\cal O}(\varepsilon^{2\ell+1})\;.
\end{split}
\end{equation}
In addition, the solution~$\psi_\nu^{\rm C}(x)$ in Eq.~(\ref{eq:psi_C_sol}) can be written in the form
\begin{align}
\psi_\nu^{\rm C}(x) = \psi_\nu^{\rm C, in}(x) + \psi_\nu^{\rm C, out}(x) \;, \label{eq:psi_C_asymptotic}
\end{align} 
where 
\begin{equation}
\begin{split}
    \psi_\nu^{\rm C, in}(x)&=e^{-i\varepsilon x}(\varepsilon x)^{\nu+1}\left(1-\frac{1}{x}\right)^{-i\varepsilon}2^\nu e^{-\pi\varepsilon}e^{i\pi(\nu+1)} \\
    &\quad \times \sum_{n=-\infty}^\infty \frac{\Gamma(g)}{\Gamma(-h+1)}a_n^\nu(-2i\varepsilon x)^n\,{}_1U_1(-h-1, g-h+1;2i\varepsilon x)\;,\\
    \psi_\nu^{\rm C, out}(x)&=e^{i\varepsilon x}(\varepsilon x)^{\nu+1}\left(1-\frac{1}{x}\right)^{-i\varepsilon}2^\nu e^{-\pi\varepsilon}e^{-i\pi(\nu+1)} \\
    &\quad \times \sum_{n=-\infty}^\infty \frac{\Gamma(g)\Gamma(g+2)}{\Gamma(-h-1)\Gamma(-h+1)}a_n^\nu(-2i\varepsilon x)^n\,{}_1U_1(g+2, g-h+1;2i\varepsilon x)\;,
\end{split}\label{Coulomb-wave-sol}
\end{equation}
with ${}_1U_1$ being the confluent hypergeometric function of the second kind.
Indeed, the form~(\ref{eq:psi_C_asymptotic}) is suitable for discussing the asymptotic behavior of the solution.\footnote{For example, in Appendix~\ref{app:reflection-MST}, we calculate the complex reflection rate for odd-parity perturbations.}
For completeness, we present the upgoing solution that satisfies the outgoing boundary condition at infinity as
\begin{align}
    \psi_\nu^{-{\rm up}}(x) &=\frac{\Gamma(\nu-1-i\varepsilon)\Gamma(\nu+1-i\varepsilon)}{\Gamma(\nu+1+i\varepsilon)\Gamma(\nu+3+i\varepsilon)}\psi^{\rm C,out}_\nu(x) \nonumber\\
    &=\left[e^{2\pi i\nu}+\frac{\sin[\pi(\nu+i\varepsilon)]}{\sin[\pi(\nu-i\varepsilon)]}\right]^{-1}\left(\frac{\sin[\pi(\nu+i\varepsilon)]}{\sin[\pi(\nu-i\varepsilon)]}(K_\nu)^{-1}\psi_\nu^{(0)}(x) - ie^{i\pi\nu}(K_{-\nu-1})^{-1}\psi_{-\nu-1}^{(0)}(x)\right)\;,
\label{MST-hor-up}
\end{align}
where in the second line we have expressed it in terms of $\psi_\nu^{(0)}(x)$ and $\psi_{-\nu-1}^{(0)}(x)$.
Since this solution is linearly independent of the solution~$\psi_\nu^{- {\rm in}}$, we will use these solutions to construct the Green's function in Section~\ref{sec:odd_sector_para}.
Note that in the solution above, when taking the low-frequency limit, the term involving $(K_{-\nu-1})^{-1}$ dominates over the other term.

\subsubsection{Tidal response for odd-parity sector}\label{ssec:tidal-MST}
Having obtained the horizon-ingoing solution~\eqref{MST-hor-in-large-x} in the previous section, we are now ready to extract the tidal response function.
From the asymptotic (i.e., large-$x$) behavior of the MST solution~\eqref{MST-hor-in-large-x}, we define the bare tidal response function as
\begin{align}
    {\cal F}_\nu^{-,\,{\rm Bare}}(\omega)\equiv \frac{\left(\text{the coefficient of $x^{-\nu}$ in $\psi_{-\nu-1}^{(0)}(x)$}\right)}{\left(\text{the coefficient of $x^{\nu+1}$ in $\psi_{\nu}^{(0)}(x)$}\right)}\;.\label{def-bare-tidal-MST}
\end{align}
It is important to point out that the presence of the renormalized angular momentum in the MST solution allows us to uniquely distinguish the growing and decaying modes at infinity at least at the first order in $\varepsilon$.
This implies that, at the linear order in frequency, one does not necessarily perform the matching with the solutions obtained from the worldline EFT (see, e.g.,~\cite{Hui:2020xxx,Charalambous:2021mea}).
Notice that this procedure is, in fact, analogous to the analytic continuation of $\ell$ to non-integer values used to resolve the ambiguity in the definition of static TLNs.
However, if one is interested in terms of second or higher order in $\varepsilon$ (e.g., dynamical TLNs), the matching with the worldline EFT would be necessary to avoid the ambiguities.
We leave this investigation to future work.

When one studies the small-$\varepsilon$ behavior of ${\cal F}_\ell^{-,\,{\rm Bare}}(\omega)$, it was pointed out in \cite{Charalambous:2021mea} that there is an ambiguity due to the pole of Gamma function contained in the MST solution, and this ambiguity can be resolved in the following manner.
First, we put $\nu=\ell+\Delta\ell$, with $\Delta\ell$ considered independent of $\varepsilon$, and define the renormalized tidal response function as
\begin{align}
    {\cal F}_\ell^{-,\,{\rm Ren}}(\omega)&=\Big[{\cal F}_\nu^{-,\,{\rm Bare}}(\omega)\exp (-2\Delta\ell \log x)\Big]_{{\rm subtract\ }(\Delta\ell)^{-1}}\;,\label{def-renormalized-response}
\end{align}
where we have subtracted terms divergent in $(\Delta\ell)^{-1}$.
The relation between $\Delta\ell$ and $\varepsilon$, which is determined by the condition~\eqref{determine-nu}, is imposed after this manipulation.
Note that this subtraction procedure corresponds to the minimal subtraction in the worldline EFT approach (see, e.g., \cite{Kol:2011vg}), and this reproduces the vanishing of static TLNs in the limit~$\varepsilon\to 0$.
(See Appendix~\ref{app:subtraction} for a more detailed discussion.)
Having said that, since the dissipative coefficient at the first order in $\varepsilon$ contains no divergent terms, TDNs are unaffected by ambiguities of the subtraction procedure.
Therefore, up to the leading order in $\varepsilon$, the tidal response function of a Schwarzschild BH can be computed as
\begin{equation}
    {\cal F}_\ell^{-,\,{\rm Ren}}(\omega)=\frac{\big[(\ell-2)!(\ell+2)!\big]^2}{(2\ell+1)!(2\ell)!}i\varepsilon+{\cal O}(\varepsilon^2)\;.
\end{equation}
From the above expression, the purely imaginary part at order $\varepsilon$ yields a set of TDNs of the Schwarzschild BH in the odd-parity sector:
\begin{align}\label{eq:TDN_Sch_GR}
    {\cal N}_{\ell,{\rm GR}}^{-}&=\frac{\big[(\ell-2)!(\ell+2)!\big]^2}{(2\ell+1)!(2\ell)!}\;.
\end{align}
Notice that the difference between our result above and those obtained in \cite{Charalambous:2021mea,HegadeKR:2024agt,Katagiri:2024wbg} is due to different normalization factors.

\paragraph{Relation to the static solution:} Let us discuss the relation between the horizon-ingoing solution obtained from the MST method and the solution of the static Regge-Wheeler equation ($\varepsilon = 0$).
Actually, at ${\cal O}(\varepsilon)$ the MST solution is essentially the same as the static solution, except that it satisfies different boundary conditions. 
(Note that the MST method ensures that the solution to the dynamical equation satisfies the ingoing behavior at the horizon.)
As we will see, the MST solution can be expressed as a linear combination of the static solutions. 

Let us consider the solutions of the static Regge-Wheeler equation (see, e.g., Eqs.~(3.14) and (3.15) of \cite{Barura:2024uog}), which are written explicitly as
\begin{align}
    \psi_{+\ell}^{\rm RW}(x) &= x^{\ell+1}\,{}_2F_1(-\ell-2, 2-\ell; -2\ell; x^{-1})\;, \quad 
    \psi_{-\ell}^{\rm RW}(x) =x^{-\ell}\,{}_2F_1(\ell-1, \ell+3; 2\ell+2; x^{-1})\;.\label{static-RW}
\end{align}
Here, $\psi_{+\ell}^{\rm RW}(x)$ is regular at $x = 1$ and $\psi_{+\ell}^{\rm RW}(x)=x^{\ell+1}[1+{\cal O}(x^{-1})]$ as $x\to\infty$, while $\psi_{-\ell}^{\rm RW}(x)$ is logarithmically divergent at $x=1$ and $\psi_{-\ell}^{\rm RW}(x)=x^{-\ell}[1+{\cal O}(x^{-1})]$ as $x\to\infty$.
It is straightforward to show that $\psi_\nu^{(0)}(x)$ in Eq.~\eqref{growing} and $ \psi_{-\nu-1}^{(0)}(x)$ in Eq.~\eqref{decaying} can be expressed as\footnote{The overall factors of ${\cal O}(\varepsilon^{-1})$ appear due to the condition~$a_0^\nu=1$ imposed just above Eq.~\eqref{eq:a_n_-nu}. One can rescale the solutions to remove these factors so that the limit~$\varepsilon\to 0$ is well-defined.}
\begin{equation}
\begin{split}\label{low-freq-MST}
    \psi_\nu^{(0)}(x)&= \frac{i(-1)^\ell}{\varepsilon} \Bigg(\frac{(2\ell)!}{\big[(\ell+2)!\big]^2}\psi_{+\ell}^{\rm RW}+\frac{\big[(\ell-2)!\big]^2}{2(2\ell+1)!\ell_2}\psi_{-\ell}^{\rm RW}\Bigg) \left[1+i\varepsilon\left(\ell_2-\frac{\ell^2+\ell+4}{2\ell(\ell+1)}\right)\right] \;,\\
    \psi_{-\nu-1}^{(0)}(x) &=\frac{i(-1)^{\ell+1}}{\varepsilon} \frac{\big[(\ell-2)!\big]^2}{2(2\ell+1)!\ell_2}\psi_{-\ell}^{\rm RW} \left[1-i\varepsilon\left(\ell_2+\frac{\ell^2+\ell+4}{2\ell(\ell+1)}\right)\right] \;,
\end{split}
\end{equation}
respectively, where we have neglected terms of higher order in $\varepsilon$.
Therefore, up to leading orders in $\varepsilon$, the horizon-ingoing solution is
\begin{equation}
    \psi_\nu^{- {\rm in}}(x) = \frac{i(-1)^{\ell}}{\varepsilon}\Bigg\{\left[1+i\varepsilon\left(\ell_2-\frac{\ell^2+\ell+4}{2\ell(\ell+1)}\right)\right]\frac{(2\ell)!}{\big[(\ell+2)!\big]^2}\psi_{+\ell}^{\rm RW}+i\varepsilon\frac{\big[(\ell-2)!\big]^2}{(2\ell+1)!}\psi_{-\ell}^{\rm RW} \Bigg\}\;.
\label{low-freq-MST-in}
\end{equation}
We see that the ratio of the coefficients of $\psi_{\pm\ell}^{\rm RW}(x)$ gives a set of TDNs, which is the same as Eq.~(\ref{eq:TDN_Sch_GR}).\footnote{Note that the ${\cal O}(\varepsilon)$ coefficient of $\psi_{+\ell}^{\rm RW}$ yields ${\cal O}(\varepsilon^2)$ corrections to the tidal response function.}
Moreover, from the expression~\eqref{low-freq-MST}, we see that TDNs do not depend on $\ell_2$ and are unaffected by subtraction procedure.

\subsection{Chandrasekhar transformation}\label{sec:even_transformation}
In the previous section, we have discussed the BH TDNs in the odd-parity sector. 
Here, we study the TDNs in the even-parity sector by using the Chandrasekhar transformation~\cite{Chandrasekhar:1975nkd,Chandrasekhar:1975zz,Chandrasekhar:1985kt}, which maps a solution of odd-parity perturbations to that of even-parity perturbations (and vice versa). 
In particular, we will show that the TLNs/TDNs we obtained in the odd-parity sector remain the same under such a transformation.

Before proceeding with the Chandrasekhar transformation, let us, for convenience, introduce the following differential operators:
\begin{equation}\label{RWZ-operator}
    {\cal H}_\ell^\pm\equiv \left(f\frac{\de}{\de r}\right)^2+\omega^2-fV_\ell^\pm\;,
\end{equation}
which allow us to write the Zerilli/Regge-Wheeler equations as $ {\cal H}_\ell^\pm \psi_\ell^\pm =0$.
The Chandrasekhar transformation is defined by the following operators:
\begin{equation}
\label{ChandraDpm}
{\cal D}_\ell^\pm\equiv f\frac{\de}{\de x}\pm \left(\frac{3f}{x(\lambda x+3)}+\frac{\lambda(\lambda+2)}{6}\right)\;.
\end{equation}
Recall that $\lambda = (\ell + 2)(\ell -1)$ and $x=r/r_g$.
Note that the operators~${\cal H}_\ell^\pm$ can be written into the form,
\begin{equation}
    {\cal H}_\ell^\pm=\frac{1}{r_g^2}\Bigg[{\cal D}_\ell^\pm {\cal D}_\ell^\mp +\left(\frac{\lambda^2(\lambda+2)^2}{36}+\varepsilon^2\right)\Bigg]\;.
\end{equation}
Then, one immediately finds that ${\cal H}_\ell^\pm{\cal D}_\ell^\pm={\cal D}_\ell^\pm{\cal H}_\ell^\mp$, and therefore $\psi_\ell^\pm\propto {\cal D}_\ell^\pm \psi_\ell^\mp$.
Note that the Chandrasekhar transformation preserves the asymptotic behaviors of the mode functions up to an overall constant factor.
Thus, in vacuum GR, the odd- and even-parity sectors share the same static TLNs and TDNs.
For the TDNs, we write
\begin{align}
   {\cal N}_{\ell,{\rm GR}}^{-}={\cal N}_{\ell,{\rm GR}}^{+} = {\cal N}_{\ell,{\rm GR}}\equiv\frac{\big[(\ell-2)!(\ell+2)!\big]^2}{(2\ell+1)!(2\ell)!}\;.
\end{align}
We will come back to this point again when analyzing the parametrized TDNs for even-parity perturbations in Section~\ref{sec:even-parity_para}.

\section{Parametrized TDN formalism}\label{sec:parametrized-formalism}
In the previous section, we have reviewed the computation of the TDNs in GR using the MST method. 
Here, we develop the parametrized formalism that allows one to compute the TDNs in a theory-agnostic manner.
Within this framework, we can also study the general structure and properties of the TDNs for a generic system.

In the parametrized formalism, for simplicity, we assume (\textit{i})~the BH background is static and spherically symmetric,\footnote{Note that a slowly rotating BH can be understood as a static and spherically symmetric BH plus small deviations associated with the BH spin.} (\textit{ii})~any deviation from vacuum GR is small, and (\textit{iii})~the system is described by a single decoupled degree of freedom.
With these assumptions, we consider a deformation of the Zerilli/Regge-Wheeler equation in the following form~\cite{Cardoso:2019mqo}:
    \begin{align}
    \left[\left(\tilde{f}\frac{{\rm d}}{{\rm d}r}\right)^2+\omega^2-\tilde{f}\tilde{V}_\ell^\pm\right]\tilde{\psi}_\ell^\pm=0\;, \label{deformed-RWZ0}
    \end{align}
where $\tilde{f}$ and $\tilde{V}_\ell^\pm$ are functions of $r$, encoding the small deviations from vacuum GR.
The horizon radius~$r=r_g$ is a simple zero of the function~$\tilde{f}$, and we assume $\tilde{f}(r)=f(r)Z(r)$, where the function~$Z$ is regular at $r=r_g$ and we recall that $f(r)=1-r_g/r$.
Note that, under our assumption, the deviation of $Z(r)$ from unity is small.
We then perform a field redefinition~$\psi_\ell^\pm\equiv \sqrt{Z}\,\tilde{\psi}_\ell^\pm$ to rewrite Eq.~\eqref{deformed-RWZ0} as
    \begin{align}
    \left[\left(f\frac{{\rm d}}{{\rm d}r}\right)^2+\frac{\omega^2}{Z^2}-f\left(V_\ell^\pm+r_g^{-2}\delta V_\ell^\pm\right)\right]\psi_\ell^\pm=0\;,
    \label{deformed-RWZ}
    \end{align}
with $\delta V_{\ell}^\pm$ denoting the deviation from the Zerilli/Regge-Wheeler potential in GR:
    \begin{align}
    r_g^{-2}\delta V_\ell^\pm=\frac{\tilde{V}_\ell^\pm}{Z}+\frac{1}{\sqrt{Z}}\frac{{\rm d}}{{\rm d}r}\left(f\frac{{\rm d}\sqrt{Z}}{{\rm d}r}\right)-V_\ell^\pm\;.
    \end{align}
Note that the factor of $r_g^{-2}$ has been inserted in front of $\delta V_\ell^\pm$ in Eq.~\eqref{deformed-RWZ}, so that $\delta V_\ell^\pm$ is dimensionless.
Since the term with $\omega^2$ is irrelevant for the TDNs, we will neglect this term.
Then, all the deviations from GR are absorbed in $\delta V_{\ell}^\pm$, which we parametrize as a series expansion in powers of $r_g/r$ as follows:\footnote{\label{footnote:notermwithj<3} Note that $\delta V_\ell^{\pm}$ in Eq.~\eqref{deformed-potential} does not involve terms with $j<3$. Otherwise, an infrared cutoff must be introduced to regularize some of the integrals in the computations that follow.}
\begin{align}\label{deformed-potential}
\delta V_\ell^{\pm}&=\sum_{j=3}^\infty \alpha_j^{\pm}\left(\frac{r_g}{r}\right)^j\;,
\end{align}
where $\alpha_j^\pm$'s are dimensionless constants and we have assumed $\delta V_\ell^\pm={\cal O}(r_g^3/r^3)$. We assume that Eq.~\eqref{deformed-potential} and the condition~$|\delta V_\ell^\pm|\ll |r_g^2V_\ell^\pm|$, which guarantees that the deviations from vacuum GR are small, hold everywhere outside the horizon ($r>r_g$), not only for $r\gg r_g$ but also including the near-horizon region. For example, if $\sum_{j=3}^{\infty}|\alpha_j^\pm|\ll 1$, then the condition~$|\delta V_\ell^\pm|\ll |r_g^2V_\ell^\pm|$ is always satisfied outside the horizon ($r>r_g$). 

We are interested in the solution to Eq.~\eqref{deformed-RWZ} satisfying the ingoing boundary condition at $r=r_g$, and we find such a solution order by order in the small deviations from GR.
Denoting the deviation from the GR solution~$\psi_\nu^{\pm {\rm in}}(r)$ by $\delta \psi_\nu^{\pm {\rm in}}(r)$, Eq.~\eqref{deformed-RWZ} can be rewritten as
\begin{align}
     \left[\left(f\frac{\de}{\de r}\right)^2-fV_\ell^{\pm}\right]\delta\psi_{\nu}^{\pm{\rm in}}\Bigg|_{\text{up to }{\cal O}(\varepsilon)}
     =f r_g^{-2} \delta V_\ell^\pm \psi_{\nu}^{\pm{\rm in}}\Bigg|_{\text{up to }{\cal O}(\varepsilon)}\;, \label{perturbed-eq}
\end{align}
where we have dropped the term with $\omega^2$ since we are interested in the solution valid only up to ${\cal O}(\varepsilon)$. Note that the solution~$\delta\psi_\nu^{\pm{\rm in}}$ satisfies the ingoing boundary condition at the horizon.
We see that Eq.~(\ref{perturbed-eq}) is an inhomogeneous equation with the source term given by the modifications to the potential and the GR solution~$ \psi_{\nu}^{\pm{\rm in}}$. 
Also, the differential operator on the left-hand side of Eq.~(\ref{perturbed-eq}) is simply the one appearing in the static Zerilli/Regge-Wheeler equation. 
Therefore, in the next subsection, we will use the Green's function method to solve Eq.~(\ref{perturbed-eq}), imposing the ingoing boundary condition at the horizon.

\subsection{Odd-parity sector}\label{sec:odd_sector_para}
Here, we are going to solve Eq.~(\ref{perturbed-eq}) for the odd-parity sector using the Green's function method.   
As pointed out before, the differential operator on the left-hand side of Eq.~(\ref{perturbed-eq}) is simply that for the static Regge-Wheeler equation in GR; therefore, we can use the MST expression for the homogeneous solutions for $\varepsilon\ll 1$.
In order to solve for the inhomogeneous solution of Eq.~(\ref{perturbed-eq}), let us introduce the Green's function~$G(x,x')$, which satisfies
\begin{align}
    \left[\frac{\de}{\de x}\left(f\frac{\de}{\de x}\right) - r_g^2V_\ell^-\right]G(x,x')=\delta(x-x')\;, \label{GreenF_def}
\end{align}
where $\delta(x)$ is the delta function and we recall that $x=r_g/r$.
By imposing the ingoing boundary condition at $x=1$ and the outgoing boundary condition at infinity, the Green's function is given by (see Appendix~A of \cite{Barura:2024uog})
\begin{align}
    G(x,y)&=\frac{\psi^{-{\rm up}}_\nu(x)\psi^{- {\rm in}}_\nu (y)\Theta(x-y)+\psi^{-{\rm up}}_\nu(y)\psi^{- {\rm in}}_\nu(x)\Theta(y-x)}{W_f\left[\psi^{-{\rm up}}_\nu(y), \psi^{- {\rm in}}_\nu(y)\right]}\;, \label{eq:green_para}
\end{align}
where $\Theta(x)$ is the Heaviside step function and we have defined the Wronskian of two functions~$F_1(x)$ and $F_2(x)$ as
\begin{align}
W_f\left[F_1(x), F_2(x)\right] \equiv f(x) \left[\frac{\de F_1(x)}{\de x} F_2(x) - F_1(x) \frac{\de F_2(x)}{\de x} \right] \;.
\end{align}
Note that we use $\psi^{-{\rm in/up}}_\nu(x)$ [Eqs.~\eqref{MST-hor-in} and \eqref{MST-hor-up}] as two linearly independent solutions.
Then, using the expressions of $\psi^{-{\rm in/up}}_\nu(x)$ in terms of the near-zone MST solutions~$\psi_{\nu}^{(0)}(x)$ and $\psi_{-\nu-1}^{(0)}(x)$ [see Eqs.~\eqref{MST-hor-in-large-x} and \eqref{MST-hor-up}],
the Green's function~(\ref{eq:green_para}) can be rewritten as
\begin{equation}\label{eq:Green_fn}
    G(x,y)=\frac{\psi_{-\nu-1}^{(0)}(x)\psi_{-\nu-1}^{(0)}(y)+\psi_\nu^{(0)}(x)\psi_{-\nu-1}^{(0)}(y)\Theta(y-x)+\psi_\nu^{(0)}(y)\psi_{-\nu-1}^{(0)}(x)\Theta(x-y)}{W_f\left[\psi^{(0)}_{-\nu-1}(y), \psi^{(0)}_\nu(y)\right]} \;,
\end{equation}
where we have neglected terms containing $K_{-\nu-1}/K_\nu \sim{\cal O}(\varepsilon^{2\ell})$.
Note that the term without the step function in Eq.~(\ref{eq:Green_fn}) is present for both $x > y$ and $y < x$.
Given the Green's function above, the solution for $\delta\psi_\nu^{\rm -in}(x)$ to Eq.~(\ref{perturbed-eq}) is
\begin{align}
    \delta\psi_\nu^{- {\rm in}}(x)&=\int_1^{\infty} \de y~ G(x,y)\psi_\nu^{- {\rm in}}(y)\delta V_\ell^-(y) \nonumber \\
    &=\frac{1}{W_f}
    \bigg\{\psi^{- {\rm in}}_\nu(x)\int_1^\infty \de y~ \psi_{-\nu-1}^{(0)}(y)\psi_\nu^{- {\rm in}}(y)\delta V_\ell^-(y) -\psi_\nu^{(0)}(x)\int_1^x\de y~\psi_{-\nu-1}^{(0)}(y)\psi_\nu^{- {\rm in}}(y)\delta V_\ell^-(y) \nonumber \\
    &\qquad\quad~+\psi_{-\nu-1}^{(0)}(x)\int_1^x\de y~\psi_\nu^{(0)}(y)\psi_\nu^{- {\rm in}}(y)\delta V_\ell^-(y) \bigg\}\;, \label{eq:GRW-inhomogeneous-sol}
\end{align}
with $W_f$ without arguments defined by $W_f\equiv W_f\left[\psi^{(0)}_{-\nu-1}(x), \psi^{(0)}_\nu(x)\right]={\rm const}$.
Note that the first term on the right-hand side reflects the boundary condition that the amplitude of the external tidal field is not affected by the deviation from GR (see footnote~\ref{footnote:notermwithj<3}).
Actually, the integrals appearing in Eq.~(\ref{eq:GRW-inhomogeneous-sol}) are similar to those computed in \cite{Barura:2024uog} in the sense that at leading order in $\varepsilon$ one replaces $\psi_\nu^{(0)}(x)$ and $\psi_{-\nu-1}^{(0)}(x)$ with the static solutions~$\psi_{+\ell}^{\rm RW}(x)$ and $\psi_{-\ell}^{\rm RW}(x)$, respectively.
Following the same procedure as in \cite{Barura:2024uog}, we obtain
\begin{equation}
\delta\psi_\nu^{- {\rm in}}(x) 
    \propto x^{\nu+1}\Bigg\{{\cal O}(x^{-1})+\delta {\cal F}_\nu^{-,\,{\rm Bare}}x^{-2\nu-1}\Big[1+{\cal O}(x^{-1})\Big]\Bigg\}+{\cal O}(\varepsilon^2
    )\;,   \label{asympt} 
\end{equation}
where the modification to the bare response function~$\delta {\cal F}_\nu^{-,\,{\rm Bare}}$ is defined by
\begin{equation}
\begin{split}
\delta {\cal F}_\nu^{-,\,{\rm Bare}}&=\Big(\delta C_+^-+\delta C_-^-\Big)\times \frac{\left(\text{the coefficient of $x^{-\nu}$ in $\psi_{-\nu-1}^{(0)}(x)$}\right)}{\left(\text{the coefficient of $x^{\nu+1}$ in $\psi_{\nu}^{(0)}(x)$}\right)}\;.  
\end{split}\label{bare-psi}  
\end{equation}
Here, the second factor can be determined from the coefficients of $\psi_{+\ell}^{\rm RW}$ and $\psi_{-\ell}^{\rm RW}$ in Eq.~\eqref{low-freq-MST}.
The coefficients~$\delta C_+^-$ and $ \delta C_-^-$ in the above expression are defined by the following integrals:
\begin{align}
    \delta C_+^- \equiv \left[\frac{1}{W_f}\int_1^x\de y\;\psi_{\nu}^{(0)}(y)\psi_\nu^{- {\rm in}}(y)\delta V_\ell^-(y)\right]_{\rm const}\;, \quad
    \delta C_-^- \equiv \frac{1}{W_f}\int_1^\infty \de x\; \psi_{-\nu-1}^{(0)}(x)\psi_\nu^{- {\rm in}}(x)\delta V_\ell^-(x)\;, \label{delta-C_(1)}
\end{align}
where the subscript~``const'' refers to the constant part of the $1/x$-expansion.
Then, the renormalized tidal response is extracted by the same formula as in the GR case [see Eq.~\eqref{def-renormalized-response}]. 

Let us now compute $\delta C_+^- + \delta C_-^-$ in Eq.~(\ref{bare-psi}).
Using Eqs.~(\ref{MST-hor-in-large-x}) and (\ref{low-freq-MST}) in (\ref{delta-C_(1)}), we find
\begin{align}
    \delta C_+^-+\delta C_-^-
    &= \frac{1}{\varepsilon^2 W_f} \left\{\int_1^x \de x\; \left[1+2i\varepsilon\left(\ell_2-\frac{\ell^2+\ell+4}{2\ell(\ell+1)}\right)\right]\left(\frac{(2\ell)!}{\big[(\ell+2)!\big]^2}\psi_{+\ell}^{\rm RW}\right)^2\delta V_\ell^-\right\}_{\rm const}  \nonumber \\
    &\quad +\frac{2i}{\varepsilon W_f} \int_1^\infty \de x\;\frac{(2\ell)!}{\big[(\ell+2)!\big]^2}\psi_{+\ell}^{\rm RW}\frac{\big[(\ell-2)!\big]^2}{(2\ell+1)!}\psi_{-\ell}^{\rm RW}\delta V_\ell^-+{\cal O}(\varepsilon^2) \;.  \label{eq:C2}
\end{align}
Note in passing that $W_f$, defined just below Eq.~\eqref{eq:GRW-inhomogeneous-sol}, scales as $\varepsilon^{-2}$.
We then define
\begin{align}
    e^{-,\,j}_{(0)}\equiv -\frac{1}{2\ell+1}\Bigg[\int_1^x \de x\; x^{-j} \left(\psi_{+\ell}^{\rm RW}\right)^2\Bigg]_{\rm const}\;,\quad e^{-,\,j}_{(1)}\equiv -\frac{2}{2\ell+1}\int_1^\infty \de x\; x^{-j} \psi_{+\ell}^{\rm RW}\psi_{-\ell}^{\rm RW}\;.\label{odd-basis}
\end{align}
It will turn out that these $e^{-,\,j}_{(0)}$ and $e^{-,\,j}_{(1)}$ serve as bases of the parametrized TLNs and TDNs, respectively.
Then, rewriting Eq.~(\ref{eq:C2}) in terms of $e^{-,\,j}_{(0)}$ and $e^{-,\,j}_{(1)}$, we have
\begin{align}
    \delta C_+^-+\delta C_-^-&=-\frac{2(2\ell)!(2\ell+1)\ell_2}{\big[(\ell+2)!\big]^2\big[(\ell-2)!\big]^2}\left[1+2i\varepsilon\frac{\ell^2+\ell+4}{2\ell(\ell+1)}+{\cal O}(\varepsilon^2)\right] \nonumber \\
    &\quad \times \sum_{j\ge 3}\alpha_j^-\bigg\{\left[1+2i\varepsilon \left(\ell_2-\frac{\ell^2+\ell+4}{2\ell(\ell+1)}\right)\right]e^{-,\,j}_{(0)}+i\varepsilon{\cal N}_{\ell}^-e^{-,\,j}_{(1)}\bigg\}\;.
\end{align}
Therefore, the deviation of the bare tidal response function is given by
\begin{align}
    \delta {\cal F}_\nu^{-,\,{\rm Bare}}&=\left[1+2i\varepsilon\frac{\ell^2+\ell+4}{2\ell(\ell+1)}+{\cal O}(\varepsilon^2)\right]\left[1-2i\varepsilon \ell_2+{\cal O}(\varepsilon^2)\right] \nonumber \\
    &\quad \times \sum_{j\ge 3}\alpha_j^-\bigg\{\left[1+2i\varepsilon \left(\ell_2-\frac{\ell^2+\ell+4}{2\ell(\ell+1)}\right)\right]e^{-,\,j}_{(0)}+i\varepsilon{\cal N}_{\ell}^-e^{-,\,j}_{(1)}\bigg\} \nonumber \\
    &=\sum_{j\ge 3}\alpha_j^-\bigg[e^{-,\,j}_{(0)}+i\varepsilon{\cal N}_{\ell}^-e^{-,\,j}_{(1)}\bigg]+{\cal O}(\varepsilon^2, (\alpha_j^-)^2)\;. \label{deviated-bare-response-odd}
\end{align}
From the above expression, the static TLNs $\mathcal{K}_\ell^-$ and the TDNs $\mathcal{N}_\ell^-$ are
\begin{align}\label{eq:TLN_TDN_para}
{\cal K}_\ell^- = 0 + \delta {\cal K}_\ell^- \;, \quad
{\cal N}_{\ell}^- = {\cal N}_{\ell,{\rm GR}} + \delta {\cal N}_{\ell}^- \;, 
\end{align}
where we have defined
\begin{align}\label{eq:para_TDNs}
    \delta {\cal K}_\ell^- \equiv \sum_{j\ge 3} \alpha_j^- e^{-,\,j}_{(0)} \;, \quad
    \delta {\cal N}_{\ell}^- \equiv {\cal N}_{\ell,{\rm GR}} \sum_{j\ge 3} \alpha_j^- e^{-,\,j}_{(1)}\;.
\end{align}
This is our main result in the parametrized TDN formalism. 
Note that the result for the static TLNs is consistent with that obtained in \cite{Katagiri:2023umb} (see also Appendix~B of \cite{Barura:2024uog}).
We see that given the coefficients~$\alpha_j^-$ and the bases~$e^{-,\,j}_{(0)}$ and $e^{-,\,j}_{(1)}$, both static TLNs and TDNs can be computed. 
It is important to point out that our method is applicable to a broad class of systems, provided that their modified Regge-Wheeler equation can be cast into the form of Eq.~(\ref{deformed-RWZ}).
We will apply our parametrized formalism to several examples in Section~\ref{sec:applications}.

Before closing this subsection, several comments are in order.
For each $\ell$, the integral that defines $e^{-,\,j}_{(0)}$ is divergent for some value(s) of $j$.
The divergent part can be isolated as a $(\Delta\ell)^{-1}$~term by shifting $\ell\to \ell+\Delta\ell$, which gives rise to a logarithmic running in the renormalized tidal response.
(For a detailed discussion on the origin of the logarithmic running, see Appendix~\ref{app:subtraction}.)
The basis~$e^{-,\,j}_{(0)}$ for some $\ell$'s was presented in Appendix~B of \cite{Barura:2024uog}.\footnote{The basis $e^{-,\,j}_{(0)}$ in our notation corresponds to $e_j$ in Appendix~B of \cite{Barura:2024uog}.}
The basis for TDNs~$e^{-,\,j}_{(1)}$ can be calculated using Eq.~(\ref{odd-basis}). 
Table~\ref{tae_e(1)} shows the explicit expressions for $e^{-,\,j}_{(1)}$ with $\ell = 2, 3, 4$ and $j \geq 3$.
For convenience, the numerical values of $e^{-,\,j}_{(1)}$ for $\ell=2,3,4$ are summarized in Table~\ref{table(1)_num_odd}.
Also, it should be pointed out that there is no divergence in $e^{-,\,j}_{(1)}$, so that the TDNs do not run logarithmically.

\begin{table}[H]
\centering
\caption{The TDN basis~$e^{-,\,j}_{(1)}$ for $\ell = 2, 3, 4$ with $j \geq 3$ in the odd-parity sector.
${\rm Di}(x) \equiv \de \log \Gamma(x)/\de x$ denotes the digamma function and $\gamma \simeq 0.5772$ is the Euler-Mascheroni constant.
The symbol~$\lim_{j\to \mathbb{Z}_{\ge 3}}$ denotes the limit where $j$ approaches an integer greater than or equal to $3$.}
\label{tae_e(1)}
\vskip2mm
\begin{tabular}{|c||c|}
\hline
$\ell$ & $e^{-,\,j}_{(1)}$ \rule[-3mm]{0mm}{8mm} \\
\hline\hline
2 & $\displaystyle \lim_{j\to \mathbb{Z}_{\ge 3}}\left\{\frac{1}{2(j-3)}+\frac{2}{3(j-4)}+\frac{1}{j-5}+\frac{2}{j-6}-\frac{2}{j-7}\left[{\rm Di}(j-7)+\gamma+\frac{1}{j-7}\right]\right\}$ \rule[-5mm]{0mm}{12mm} \\
\hline
3 & \begin{tabular}{rl}$\displaystyle \lim_{j\to \mathbb{Z}_{\ge 3}}\bigg\{\hskip-4mm$&$\displaystyle \frac{1}{2(j-3)}+\frac{16}{15(j-4)}+\frac{3}{j-5}+\frac{14}{j-6}-\frac{120}{j-8}+\frac{108}{j-9}$ \rule[-4mm]{0mm}{11mm} \\
&$\displaystyle -\frac{2(36-5j+j^2)}{(j-7)(j-8)(j-9)}\left[{\rm Di}(j-7)+\gamma+\frac{1}{j-7}\right]\bigg\}$ \rule[-5mm]{0mm}{12mm}
\end{tabular}\\
\hline
4 & \begin{tabular}{rl}$\displaystyle \lim_{j\to \mathbb{Z}_{\ge 3}}\bigg\{\hskip-4mm$&$\displaystyle \frac{1}{2(j-3)}+\frac{8}{5(j-4)}+\frac{113}{15(j-5)}+\frac{318}{5(j-6)}-\frac{2520}{j-8}+\frac{7812}{j-9}-\frac{8624}{j-10}+\frac{9800}{3(j-11)}$ \rule[-4mm]{0mm}{11mm} \\
&$\displaystyle -\frac{2(2712-890j+203j^2-10j^3+j^4)}{(j-7)(j-8)(j-9)(j-10)(j-11)}\left[{\rm Di}(j-7)+\gamma+\frac{1}{j-7}\right]\bigg\}$ \rule[-5mm]{0mm}{12mm}
\end{tabular}\\
\hline
\end{tabular}
\end{table}

\begin{table}[H]
    \centering
    \caption{Numerical values of the TDN basis~$e^{-,\,j}_{(1)}$ for $\ell=2, 3, 4$ with $3 \leq  j \leq 9 $ in the odd-parity sector, derived from the formulae in Table~\ref{tae_e(1)}.}
    \label{table(1)_num_odd}
    \vskip2mm
    \begin{tabular}{|c||c|c|c|}
    \hline
    $j$ &  $e^{-,\,j}_{(1)}\big|_{\ell=2}$ & $e^{-,\,j}_{(1)}\big|_{\ell=3}$ & $e^{-,\,j}_{(1)}\big|_{\ell=4}$ \rule[-3mm]{0mm}{8mm}\\
    \hline\hline
    $3$ & $-1.041666666667$ & $-0.641666666667$ & $-0.475000000000$\\ \hline
    $4$ & $-0.722222222222$ & $-0.435555555556$ & $-0.320000000000$\\ \hline
    $5$ & $-0.583333333333$ & $-0.350000000000$ & $-0.256666666667$\\ \hline
    $6$ & $-0.500000000000$ & $-0.300000000000$ & $-0.220000000000$\\ \hline
    $7$ & $-0.442645911474$ & $-0.266147786856$ & $-0.195330081702$\\ \hline
    $8$ & $-0.400000000000$ & $-0.241245311546$ & $-0.177262653580$\\ \hline
    $9$ & $-0.366666666667$ & $-0.221919479739$ & $-0.163286812230$\\ \hline
    \end{tabular}
\end{table}

Furthermore, one can show that the basis~$e_{(1)}^{-,\,j}$ satisfies the following recurrence relation:
    \begin{align}
    (j-5)(j-1)(j+3)e_{(1)}^{-,\,j+2}+(2j-3)\left[6+2\ell(\ell+1)-j(j-3)\right]e_{(1)}^{-,\,j+1}& \nonumber \\
    +(j-2)(j-2\ell-3)(j+2\ell-1)e_{(1)}^{-,\,j}&=0 \quad
    (j\ge 3)\;,
    \end{align}
which comes from the fact that the TDNs are invariant under a redefinition of the variable.
(See \cite{Kimura:2020mrh} for details in the context of parametrized BH quasinormal ringdown formalism.
See also \cite{Katagiri:2023umb} for an application of such a recurrence relation to parametrized TLNs.)
This recurrence relation yields $e_{(1)}^{-,\,j+1}/e_{(1)}^{-,\,j}=1-j^{-1}+{\cal O}(j^{-2})$ for $j\gg 1$, implying that $e_{(1)}^{-,\,j}\sim {\cal O}(j^{-1})$.
Therefore, provided that the series expansion~\eqref{deformed-potential} of $\delta V_\ell^-$ converges, the series~$\sum_{j\ge 3} \alpha_j^- e^{-,\,j}_{(1)}$ also converges.

\subsection{Even-parity sector}\label{sec:even-parity_para}
The procedure applied to the odd-parity sector can be straightforwardly extended to the even-parity sector.\footnote{Instead of directly working on the Zerilli equation with a corrected potential, one may alternatively recast the equation into the Regge-Wheeler form with a corresponding correction by use of the Chandrasekhar transformation.}
However, thanks to the Chandrasekhar transformation, which does not change the separations between the source and the induced response (see Section~\ref{sec:even_transformation} for detailed discussion), we can directly obtain the even-parity homogeneous solutions from the odd-parity ones. Remember that all non-GR modifications are included in the source term in our computation.
Therefore, the homogeneous solution~$\psi_\ell^{+{\rm in}}(x)$, which is of even-parity and ingoing at the horizon, is obtained by $\psi_\ell^{+{\rm in}}(x) \propto {\cal D}_\ell^+ \psi_\ell^{-{\rm in}}(x)$.
Since the solution $\psi_\ell^{-{\rm in}}(x)$ can be written in terms of the static Regge-Wheeler solutions $\psi_{\pm\ell}^{\rm RW}(x)$ [see Eq.~(\ref{static-RW})], in practice we perform the replacement $\psi_{\pm\ell}^{\rm RW} \to \psi_{\pm \ell}^{\rm Z} \propto {\cal D}_\ell^+\psi_{\pm\ell}^{\rm RW}$ in Eq.~(\ref{odd-basis}) to obtain the bases~$e^{+,\,j}_{(0)}$ and $e^{+,\,j}_{(1)}$ for the even-parity sector.
For example, for $\ell=2$, we have
\begin{equation}
\begin{aligned}
    \psi_{+2}^{\rm Z}(x) &=\frac{x \left(4 x^3+6 x^2-3\right)}{4 x+3}\;, \\ 
    \psi_{-2}^{\rm Z}(x) &=-\frac{5 x \big[3 \left(4 x^3+6 x^2-3\right) \log \left(\frac{x-1}{x}\right)+12 x^2+24 x+13\big]}{12 x+9}\;,
\end{aligned}
\end{equation}
where the overall factors have been chosen so that $\psi_{+2}^{\rm Z}(x)=x^3[1+{\cal O}(x^{-1})]$ and $\psi_{-2}^{\rm Z}(x)=x^{-2}[1+{\cal O}(x^{-1})]$ as $x\to \infty$.
Then, the bases for the deviations of static TLNs and TDNs are given by
\begin{align}
    e^{+,\,j}_{(0)}\equiv -\frac{1}{2\ell+1}\bigg[\int_1^x \de x\; x^{-j} \left(\psi_{+\ell}^{\rm Z}\right)^2\bigg]_{\rm const}\;,\quad e^{+,\,j}_{(1)}\equiv -\frac{2}{2\ell+1}\int_1^\infty \de x\; x^{-j} \psi_{+\ell}^{\rm Z}\psi_{-\ell}^{\rm Z}\;.\label{even-basis}
\end{align}
Since the analytic expressions of these bases are complicated, we directly obtain their numerical values. 
In Table~\ref{table(1)_even}, we summarize the numerical values of $e^{+,\,j}_{(1)}$ for $\ell = 2, 3, 4$ with $3 \leq j \leq 9$.
For completeness, we express the static TLNs and the TDNs for the even-parity sector as
\begin{align}\label{eq:TLN_TDN_even_para}
{\cal K}_\ell^+ = 0 + \delta {\cal K}_\ell^+ \;, \quad
{\cal N}_{\ell}^+ = {\cal N}_{\ell,{\rm GR}} + \delta {\cal N}_{\ell}^+ \;, 
\end{align}
where we have defined
\begin{align}\label{eq:para_TDNs_even}
\delta {\cal K}_\ell^+ \equiv \sum_{j\ge 3} \alpha_j^+ e^{+,\,j}_{(0)}  \;, \quad
\delta {\cal N}_{\ell}^+ \equiv {\cal N}_{\ell,{\rm GR}} \sum_{j\ge 3} \alpha_j^+ e^{+,\,j}_{(1)} \;.
\end{align}
For $\ell=2,3,4$, the numerical values of $e^{+,\,j}_{(0)}$ for $j\ge 2\ell+4$ coincide with the results in \cite{Katagiri:2023umb} by multiplying the normalization factor for each $\ell$.\footnote{The mismatch at small $j$ is explained in Appendix~B of \cite{Barura:2024uog} for the odd-parity case.
In addition, one should take care of the logarithmic dependence~$\log \left(r/r_g\right)$ for $j\le 2\ell+3$ for the basis of the TLNs.}
We note that for positive $\alpha_j^+$ (i.e., when the effective potential receives positive corrections), the TDNs receive negative corrections since the basis~$e^{+,\,j}_{(1)}$ in Table~\ref{table(1)_even} are all negative [see Eq.~(\ref{eq:para_TDNs_even})].
Hence, in this particular case, the TDNs are smaller than the GR value.
\begin{table}[H]
    \centering
    \caption{Numerical values of the TDN basis~$e^{+,\,j}_{(1)}$ for $\ell=2, 3, 4$ with $3 \leq  j \leq 9 $ in the even-parity sector.}
    \label{table(1)_even}
    \vskip2mm
    \begin{tabular}{|c||c|c|c|}
    \hline
    $j$ &  $e^{+,\,j}_{(1)}\big|_{\ell=2}$ & $e^{+,\,j}_{(1)}\big|_{\ell=3}$ & $e^{+,\,j}_{(1)}\big|_{\ell=4}$ \rule[-3mm]{0mm}{8mm}\\
    \hline\hline
    $3$ & $-1.004006721789$ & $-0.637111262697$ & $-0.473898412947$\\ \hline
    $4$ & $-0.683777342020$ & $-0.430647317440$ & $-0.318785717845$\\ \hline
    $5$ & $-0.547035195906$ & $-0.345233772428$ & $-0.255473831298$\\ \hline
    $6$ & $-0.466233871050$ & $-0.295476454331$ & $-0.218858389699$\\ \hline
    $7$ & $-0.411267789636$ & $-0.261875994989$ & $-0.194244542867$\\ \hline
    $8$ & $-0.370759429190$ & $-0.237209848515$ & $-0.176230970170$\\ \hline
    $9$ & $-0.339313738762$ & $-0.218099208073$ & $-0.162304829126$\\ \hline
    \end{tabular}
\end{table}

Similarly to the case of odd-parity perturbations, one can obtain a recurrence relation that the basis~$e^{+,\,j}_{(1)}$ satisfies.
Written explicitly,
    \begin{align}
    27(j+2)^3e_{(1)}^{+,\,j+5}
    +9\left[3(j^3+3)(\lambda-2)+j(9j+13)(\lambda-3)\right]e_{(1)}^{+,\,j+4}& \nonumber \\
    +9\left[j^3(\lambda^2-6\lambda+3)-9j^2(\lambda-1)+j(4\lambda^2-7\lambda+6)+2\lambda(\lambda-2)\right]e_{(1)}^{+,\,j+3}& \nonumber \\
    +\lambda \left[j^3(\lambda^2-18\lambda+27)-3j^2\lambda(\lambda-9)+j(23\lambda^2-21\lambda-27)-3\lambda(\lambda-4)\right]e_{(1)}^{+,\,j+2}& \nonumber \\
    +\lambda^2 \left[-j^3(2\lambda-9)+9j^2(\lambda-3)+j(4\lambda^2-25\lambda-6)-6\lambda(\lambda-1)\right]e_{(1)}^{+,\,j+1}& \nonumber \\
    +\lambda^3(j-2)(j^2-4j-4\lambda-5)e_{(1)}^{+,\,j}&=0 \quad
    (j\ge 3)\;.
    \end{align}
Again, the recurrence relation yields $e_{(1)}^{+,\,j+1}/e_{(1)}^{+,\,j}=1-j^{-1}+{\cal O}(j^{-2})$ for $j\gg 1$, implying that $e_{(1)}^{+,\,j}\sim {\cal O}(j^{-1})$.
Therefore, provided that the series expansion~\eqref{deformed-potential} of $\delta V_\ell^+$ converges, the series~$\sum_{j\ge 3} \alpha_j^+ e^{+,\,j}_{(1)}$ also converges.

\section{Applications of parametrized TDN formalism}\label{sec:applications}
In this section, we apply our parametrized TDN formalism developed in the previous section to three examples, where the linear perturbation equation can be recast into the form of Eq.~\eqref{deformed-RWZ}.

\subsection{EFT with timelike scalar profile}\label{sec:EFT_gen_RW}
In this subsection, we compute TDNs using our parametrized formalism in the context of the EFT of scalar-tensor gravity with timelike scalar profile~\cite{Mukohyama:2022enj,Mukohyama:2022skk}.
As we shall review in Appendix~\ref{app:EFT-BHPT}, the generalized Regge-Wheeler equation~\eqref{gRW} can be derived on a generic static and spherically symmetric BH background, to which our parametrized TDN formalism applies.
For a stealth Schwarzschild background, one can show that the master equation reduces to the Regge-Wheeler equation in GR upon a rescaling of parameters~\cite{Mukohyama:2023xyf,Nakashi:2023vul}, which allows us to obtain the TDNs by a simple rescaling of the GR values (see Appendix~\ref{app:stealth}).
Therefore, here we focus on a non-stealth background and consider the Hayward BH as a concrete example, whose metric is given as follows:
    \begin{align}
    \bar{g}_{\mu\nu}\de x^\mu \de x^\nu 
    = -A(r)\de t^2 + \frac{\de r^2}{A(r)} + r^2(\de\theta^2 + \sin^2\theta\,\de\varphi^2)\;, \quad
    A(r)=1-\frac{r_s r^2}{r^3+\sigma^3}\;,
    \end{align}
where $r_s$ and $\sigma$ are constants of length dimension, with $\sigma$ characterizing the deviation from the Schwarzschild metric.
Note that the metric approaches the Schwarzschild metric as $r\to\infty$, while it possesses a de Sitter core at $r=0$ in the case of $\sigma>0$.
As mentioned in Appendix~\ref{app:gRW}, the function~$F(r)$ defined in Eq.~\eqref{gRW_building_blocks} fixes the position of the horizon~$r=r_g$ for the odd-parity GWs.
At the first order in $\eta\equiv \sigma^3/r_g^3$, the correction to the effective potential is given by~\cite{Barura:2024uog}\footnote{In obtaining the expression of the effective potential, the integration constant~$C$ in Eq.~\eqref{sol_alpha} has been chosen so that $C+M_\star^2 r_s=0$. With this choice, one can show that the propagation speed of the odd modes approaches that of light at spatial infinity.}
    \begin{align}
    \delta V_\ell^-=\left[-\frac{\ell(\ell+1)(x^2+x+1)}{x^5}+\frac{8x^3+9x^2-26x+21}{2x^6}\right]\eta\;, \label{V_Hayward_eta_exp}
\end{align}
and a comparison with Eq.~(\ref{deformed-potential}) yields
\begin{align}
\alpha_3^- = \left[- \ell(\ell + 1) + 4\right]\eta \;, \quad
\alpha_4^- = \left[- \ell(\ell + 1) + \frac{9}{2}\right]\eta \;, \quad
\alpha_5^- = \left[- \ell(\ell + 1) - 13\right]\eta \;, \quad
\alpha_6^- = \frac{21}{2}\eta \;.
\end{align}

We are now ready to apply the formula~\eqref{eq:TLN_TDN_para} with the list of the basis~$e_{(1)}^{-,\,j}$ in Table~\ref{tae_e(1)} to compute the correction to the TDNs.
For the case of $\ell=2$, we obtain
\begin{align}\label{eq:TDN_ell=2_Hayward}
    \delta {\cal N}_2^-&=\frac{9}{5}\eta\;,
\end{align}
which can be verified through a direct calculation of \eqref{bare-psi} (see Appendix~\ref{app:MST-specific}).
Similarly, for $\ell=3$ and $4$, we obtain
\begin{align}
    \delta {\cal N}_3^-=\frac{1}{18}\eta\;, \quad  \delta {\cal N}_4^-=\frac{13}{4900}\eta\;,
\end{align}
respectively.

\subsection{Einstein-Maxwell system}\label{subsection:RN}
Here, we apply our parametrized formalism to compute the TDNs for the odd-parity perturbations on a Reissner-Nordstr\"om BH in the Einstein-Maxwell system with small electric charge.
The background metric is given by
    \begin{align}
    \bar{g}_{\mu \nu} \de x^\mu \de x^\nu=-A_{\rm RN}(r)\de t^2+\frac{\de r^2}{A_{\rm RN}(r)}+r^2(\de\theta^2+\sin^2\theta\,\de\varphi^2) \;,
    \end{align}
where
    \begin{align}
    A_{\rm RN}(r)=1-\frac{2M}{r}+\frac{Q^2}{r^2}\;,
    \end{align}
with $M$ and $Q$ being parameters of length dimension, respectively corresponding to the mass and charge of the BH.
Provided that $M^2>Q^2>0$, the metric has two distinct horizons determined by $A_{\rm RN}(r)=0$, i.e., $r=r_\pm\equiv M\pm\sqrt{M^2-Q^2}$.
In what follows, the outer horizon radius~$r_+$ is identified as $r_g$ used in our parametrized TDN formalism.
When $|Q/M|\ll 1$, we find that the quantity
    \begin{align}
    \eta\equiv \frac{r_-}{r_g}=\frac{Q^2}{4M^2}+{\cal O}\left(\frac{Q^4}{M^4}\right)\;,
    \end{align}
is small, which controls the deviation from the vacuum GR.

In the odd-parity sector of the Einstein-Maxwell system, the metric and gauge field perturbations constitute two physical degrees of freedom, whose corresponding master equations can be decoupled by a suitable choice of master variables.
The master equation for the mode corresponding to the odd-parity metric perturbations in the limit~$Q\to 0$ can be written as follows~\cite{Moncrief:1974gw,Chandrasekhar:1985kt}:
    \begin{align}
    \left[\left(A_{\rm RN}\frac{{\rm d}}{{\rm d}r}\right)^2+\omega^2-A_{\rm RN}V_{{\rm RN}, \ell}^-\right]\Psi_{{\rm RN}, \ell}^-=0\;,
    \end{align}
where
    \begin{align}
    V_{{\rm RN}, \ell}^-(r)=\frac{\ell(\ell+1)}{r^2}+\frac{4\eta r_g^2}{r^4}-\frac{3r_g(1+\eta)}{2r^3}\left[1+\sqrt{1+\frac{16\eta(\ell-1)(\ell+2)}{9(1+\eta)^2}}\right]\;.
    \end{align}
The above equation reduces to the negative-sign branch of Eq.~\eqref{deformed-RWZ} via the following field redefinition:
\begin{equation}
\label{PhiRNm}
\psi_\ell^-=\sqrt{Z}\,\Psi_{{\rm RN},\ell}^-\;, \quad
Z=\frac{A_{\rm RN}}{f}
=1-\frac{r_-}{r}\;,
\end{equation}
where we recall that $f=1-r_g/r$.
In this case, the correction to the effective potential is given by
\begin{equation}\label{eq:delta_V_RN}
\delta V_\ell^{-} = \alpha_3^{-}\left(\frac{r_g}{r}\right)^3+\alpha_4^{-}\left(\frac{r_g}{r}\right)^4+{\cal O}\left(\eta^2\right) \;,
\end{equation}
with the coefficients~$\alpha_i^-$ given by
\begin{equation}
\label{RNalpham}
\alpha_3^{-}=-\frac{\ell(\ell+1)+4}{3} \eta \;, \quad \alpha_4^{-}=\frac{5}{2}\eta \;.
\end{equation}
Using the expression for the basis~$e^{-,\,j}_{(1)}$ (see Table~\ref{tae_e(1)}) and Eq.~(\ref{RNalpham}) in Eq.~(\ref{eq:para_TDNs}), we obtain
\begin{align}
    \delta{\cal N}_2^-=\frac{1}{3}\eta \;,\quad \delta{\cal N}_3^-=\frac{1}{108}\eta \;, \quad  \delta{\cal N}_4^-=\frac{1}{2352}\eta \;,
\end{align}
for $\ell=2,3,4$, respectively.

Let us comment on the discrepancy between our results and those obtained in \cite{Katagiri:2024fpn}, which is the same as Appendix~B of \cite{Barura:2024uog} in the case of static TLNs.
Note that in our formalism based on the MST method, the coefficient of $x^{-\ell}$ in the first-order solution does not directly correspond to the TDNs.
In our approach, we extract the TDNs using the analytic continuation of $\ell$, which is automatically implemented by the renormalized angular momentum~$\nu$. 
This procedure does not change the total solution itself, but unambiguously separates the source and response contributions.
Also, this prescription guarantees that the TDNs we defined remain invariant under redefinition of master variables (see discussion below Eq.~(3.33) of \cite{Barura:2024uog}).
Physically, this involves removing the PN correction for the metric sourced by a point particle and isolating the contribution solely from finite-size effects as TDNs (see, e.g., \cite{Creci:2021rkz,Ivanov:2022hlo}).

\subsection{Higher-curvature extension of GR}
We now apply our parametrized formalism to compute the TDNs in the EFT extension of GR by higher-order curvature corrections~\cite{Endlich:2017tqa,Cardoso:2018ptl,deRham:2020ejn,Cano:2021myl}. 
For simplicity, we take into account only the parity-preserving and dimension-8 operators as in \cite{Endlich:2017tqa,Cardoso:2018ptl,Katagiri:2023umb}.\footnote{Note that this EFT respects 4d diffeomorphism invariance, as opposed to the EFT of scalar-tensor gravity in the unitary gauge discussed in Subsection~\ref{sec:EFT_gen_RW}.
Also, the reason for starting from 8-dimensional operators is given in \cite{Camanho:2014apa} (see also Section~2.2 of \cite{Endlich:2017tqa}) as follows: the causality constraint suggests that 6-dimensional operators require an ultraviolet completion involving an infinite tower of higher-spin particles coupled to (standard model) matter fields.
However, we have not observed any additional long-range forces in the sub-kilometer distance scale of $\tilde{\Lambda}^{-1}$; thus, 6-dimensional operators should be suppressed by a much higher energy scale.
In fact, it remains an open question whether the same result applies to other higher-curvature terms or not.
Here, we only focus on the 8-dimensional operators as a working example.}
In this case, the action is given by
\begin{align}
    S_{\mathrm{eff}} & =\int\de^{4} x \frac{\sqrt{-g}}{16\pi}\left(R+\frac{c_1}{\tilde{\Lambda}^6}\mathcal{C}^{2}+\frac{c_2}{\tilde{\Lambda}^6}\tilde{\mathcal{C}}^{2}\right)\;, \label{8-dim-action} 
\end{align}
where
\begin{align}
\mathcal{C} & \equiv R_{\alpha \beta \gamma \delta} R^{\alpha \beta \gamma \delta}, \quad \tilde{\mathcal{C}} \equiv R_{\alpha \beta \gamma \delta} {}^*\!R^{\alpha \beta \gamma \delta} \;.
\end{align}
Here, the dual of the Riemann tensor has been defined by ${}^*\!R^{\alpha \beta \gamma \delta}\equiv \epsilon^{\alpha\beta}{}_{\mu\nu}R^{\mu\nu\gamma\delta}$, with $\epsilon_{\alpha\beta\gamma\delta}$ being the totally antisymmetric tensor defined so that $\epsilon_{0123}=\sqrt{-g}$.
The parameter~$\tilde{\Lambda}$ denotes the cutoff of the EFT, and $c_i$'s are dimensionless coefficients.\footnote{In the present case, we choose $c_1>0$ and $c_2>0$.
This choice is consistent with the positivity bounds derived in \cite{Bellazzini:2015cra,deRham:2019ctd}.}

The magnitude of higher-curvature corrections depends on the curvature scale of the spacetime. 
In the case of a static and spherically symmetric BH, the curvature scale around the horizon is set by the BH mass~$M$.
Let us introduce the dimensionless couplings as
\begin{align}
   \eta_i\equiv \frac{c_i}{(M\tilde{\Lambda})^6}\;,
\end{align}
for $i=1,2$.
Within the validity of the EFT, we require that $|\eta_{i}|\ll 1$.
In this way, in the presence of the parameters~$\eta_i$, the background metric is deviated from that of the Schwarzschild spacetime with mass~$M$.
Up to the first order in $\eta_i$'s, the deformed background solution is given by~\cite{Cardoso:2018ptl,Cano:2021myl}
    \begin{align}\label{eq:sol_highercurv}
    \bar{g}_{\mu \nu} \de x^\mu \de x^\nu=-N_{\rm EFT}^2B_{\rm EFT}\de t^2+\frac{\de r^2}{B_{\rm EFT}}+r^2(\de\theta^2+\sin^2\theta\,\de\varphi^2) \;,
    \end{align}
where\footnote{If one takes into account 10-dimensional operators in the effective action~\eqref{8-dim-action}, the metric functions receive corrections of order $(M\tilde{\Lambda})^{-8}$.}
    \begin{align}
    \begin{split}
    N_{\rm EFT}(r)&=1-\eta_1\frac{1792M^9}{r^9}+{\cal O}\left((M\tilde{\Lambda})^{-12}\right) \;, \\
    B_{\rm EFT}(r)&=1-\frac{2M}{r}+\eta_1\left(\frac{4608M^9}{r^9}-\frac{8576M^{10}}{r^{10}}\right)+{\cal O}\left((M\tilde{\Lambda})^{-12}\right) \;.
    \end{split}\label{eq:corr_high_curv}
    \end{align}
The radius of the trapping horizon is determined by $B_{\rm EFT}(r)=0$, which yields
    \begin{align}
    \frac{r_g}{2M}=1-\frac{5}{8}\eta_1+{\cal O}\left((M\tilde{\Lambda})^{-12}\right)\;,
    \end{align}
and it is a black hole horizon as $0<N_{\rm EFT}(r)<\infty$ for $r\geq r_g$. 

In what follows, we study the corrections to the TDNs arising from $\eta_1$ and $\eta_2$ separately.

\subsubsection{\texorpdfstring{$\eta_2$ correction}{eta2 correction}}
We first consider the case where $\eta_1=0$ and $\eta_2\neq 0$.
In this case, looking at Eq.~(\ref{eq:corr_high_curv}), we see that the background solution is the same as the Schwarzschild metric at least at first order in the higher-curvature corrections.
At the first order in $\eta_2$, the correction to the Regge-Wheeler potential is given by~\cite{Cardoso:2018ptl}
\begin{equation}
\delta V_\ell^{-} =-18\eta_2\left(\ell+2\right)\left(\ell+1\right)\ell\left(\ell-1\right)\left(\frac{r_g}{r}\right)^{10} \;,
\end{equation}
which gives 
\begin{align}
\alpha_{10}^- =  -18\eta_2\left(\ell+2\right)\left(\ell+1\right)\ell\left(\ell-1\right) \;.
\end{align}
Using Eq.~(\ref{eq:para_TDNs}) and Table~\ref{tae_e(1)}, we obtain
\begin{align}
    \delta{\cal N}_2^-=\frac{5136}{175}\eta_2\;,\quad \delta{\cal N}_3^-=\frac{260}{147}\eta_2\;,\quad  \delta{\cal N}_4^-= 0.1396 \eta_2\;,
\end{align}
for $\ell=2,3,4$, respectively.

Let us comment on the even-parity perturbations.
Since the operator~$\tilde{C}^2$ in Eq.~\eqref{8-dim-action} does not contribute to the quadratic action of even-parity perturbations, the master equation remains the same as in GR.
Therefore, in this case the TDNs for the even-parity perturbations are the same as those in GR.

\subsubsection{\texorpdfstring{$\eta_1$ correction}{eta1 correction}}
Let us consider the case with $\eta_1 \neq 0$ and $\eta_2 = 0$.
From Eq.~(\ref{eq:corr_high_curv}), we see that the background metric slightly deviates from the Schwarzschild background.
For the odd-parity sector, one finds that the corrections to the Regge-Wheeler potential are given by \cite{Cardoso:2018ptl}
\begin{equation}
\label{deltaVepsilon1}
\delta V_\ell^-=\sum_{j=3}^{12}\alpha_j^-(\ell,\,\eta_1)\left(\frac{r_g}{r}\right)^j\;,
\end{equation}
where the coefficients~$\alpha_j^-$ are proportional to $\eta_1$.
Here, we do not show the full expression of $\alpha_j^-$, but the formula for $\ell =2$ can be found in \cite{Katagiri:2023umb}.\footnote{Note that $\delta V_\ell^-$, $r_{\rm H}$, and $\epsilon_1$ therein should be understood, respectively as $r_g^{-2}\delta V_\ell^-$, $r_g$, and $-\eta_1$ in our notation.}
Following the same calculation as before, one can straightforwardly obtain the corrections to the TDN for $\ell = 2$ as
\begin{align}
    \delta{\cal N}_{\ell=2}^-= \mathcal{N}_{\ell,{\rm GR}} \sum_{j=3}^{12}\alpha_j^- e_{(1)}^{-,\,j}\Bigg|_{\ell=2}=-\frac{2719}{7000}\eta_1\;.
\end{align}
We see that the $\eta_1$~correction decreases the TDN for $\ell=2$ modes, while it reduces the horizon radius and increases the effective potential for $r>r_g$.
It is also straightforward to apply our parametrized formalism to compute the TDNs for the even-parity perturbations in this case by use of the expression of $\alpha_j^+$ in \cite{Cardoso:2018ptl}.

\section{Conclusions}\label{sec:conclusions}
In this paper, we have developed a theory-agnostic parametrized formalism to compute the tidal dissipation numbers (TDNs) of a non-rotating black hole (BH) in generic theories, for which the perturbations can be described by a master equation close to that in the vacuum solution in general relativity (GR).
The framework presented in this work assumes (\textit{i})~a static and spherically symmetric background spacetime, (\textit{ii})~small deviations of the perturbation equations from the ones in GR, i.e., the Regge-Wheeler/Zerilli equations, and (\textit{iii})~the absence of couplings among different physical degrees of freedom.
By using this formalism, one can quantify the first-order deviation of the BH TDNs from the values in GR.
In addition, given master equations in the form of Eq.~\eqref{deformed-RWZ}, one can straightforwardly compute the corresponding TDNs using Eq.~\eqref{eq:para_TDNs}.

In Section~\ref{sec:review-tidal-response}, we have reviewed the Mano-Suzuki-Takasugi (MST) method to deal with the horizon-ingoing solution in 4d GR. 
In the MST method, there exists a set of renormalized angular momentum~$\nu=\ell+\ell_2(r_g\omega)^2+{\cal O}((r_g\omega)^4)$ with $r_g$ being the horizon radius, which allows us to consistently resolve the issue of the source/response separations.\footnote{It is worth mentioning that the higher-order corrections of this renormalized angular momentum are equivalent to the scaling of the dynamical multipole moments when resumming the higher-order tail in the worldline EFT~\cite{Goldberger:2009qd,Porto:2016pyg,Almeida:2021jyt,Ivanov:2025ozg}.}
With that procedure, we have defined the tidal response function as a ratio of the coefficient of the decaying mode in $\psi^{(0)}_{-\nu-1}(x)$ to that of the growing mode in $\psi^{(0)}_\nu(x)$ without ambiguities at least at first order in the frequency.
Indeed, this can be achieved by the matching between MST horizon-ingoing solution of BH perturbation theories and the perturbations sourced by the worldline EFT~\cite{Chakrabarti:2013lua}. Moreover, in Section~\ref{sec:even_transformation}, we have discussed the Chandrasekhar transformation which generates an opposite-parity solution from a given one. With this transformation, we have concluded that the TDNs for both parity sectors are identical in GR. 

In Section~\ref{sec:parametrized-formalism}, we have established the parametrized BH TDN formalism where the deviations from the Regge-Wheeler/Zerilli equations are treated perturbatively.
Focusing on the odd-parity sector, we have obtained the first-order solution~$\delta \psi_\nu^{-{\rm in}}$ using the Green's function method.   
From such a solution, we have defined the modification to the bare response function~$\delta {\cal F}_\nu^{-,\,{\rm Bare}}$ and extracted both the static TLNs and the TDNs, which is our main result in this paper [see Eqs.~\eqref{deviated-bare-response-odd}--\eqref{eq:para_TDNs}].
The result for the even-parity sector~\eqref{eq:para_TDNs_even} can be obtained by applying the Chandrasekhar transformation.
For convenience, in Tables~\ref{tae_e(1)}--\ref{table(1)_even}, we have presented the analytic expressions and/or the numerical values of the bases~$e_{(1)}^{-,\,j}$ and $e_{(1)}^{+,\,j}$ for $\ell = 2,3,4$ and $j \geq 3$.
We note that our parametrized formula explicitly shows that the divergent part of the bare TDNs is absent even in setups beyond GR.
This implies that the logarithmic running of TDNs does not occur at this level.
(See Appendix~\ref{app:subtraction} for the origin of the logarithmic running for the TLNs, which correspond to a conserved response.)

In Section~\ref{sec:applications}, we have applied our parametrized formalism to calculate the TDNs for three examples.
In the first example, we have focused on the generalized Regge-Wheeler potential, derived from the EFT of BH perturbations with timelike scalar profile~\cite{Mukohyama:2022enj,Mukohyama:2022skk}.
In particular, we have computed the corrections to the GR TDNs when the background metric is assumed to be the Hayward background.
In the second example, we have calculated the TDNs for the odd-parity perturbations on the Reissner-Nordstr\"om background in the Einstein-Maxwell system at the leading order in the electric charge.
Furthermore, in the last example, we have evaluated the corrections to the GR TDNs in the presence of higher-order curvature terms.

There are several future directions that we would like to explore.
First, an obvious extension of this work is to generalize our parametrized formalism to a stationary and axisymmetric BH background.\footnote{In fact, this can be straightforwardly generalized to a rotating background since the MST solution of the Teukolsky equation was found in \cite{Mano:1996gn,Mano:1996vt,Sasaki:2003xr}. Therefore, with that solution, one can, in principle, formulate the parametrized formalism for the modified Teukolsky equation (see Ref.~\cite{Cano:2025zyk} for the corrected TLNs of non-rotating BHs in this context).}
By doing so, it would generally capture the effects of BH spin in both the static TLNs and the TDNs. 
Second, it is interesting to extend our formalism to systems where there are couplings among different physical degrees of freedom.
This study would reveal a rich structure of the tidal response in the alternative theories of gravity.
Third, it would be nice to formulate the parametrized formalism taking into account the higher-order corrections in frequency in the tidal response function.
This would require the matching with the solutions obtained from the worldline EFT.\footnote{Indeed, clarifying the connection with the point-particle EFT would also resolve the relation between TLNs and Green's functions in BH perturbations~\cite{DeLuca:2024ufn}.} 
We leave this investigation to our future work~\cite{Kobayashi:2025}.
Finally, with the increasing number of future GW observations, it is important to study the impact of tidal dissipation on the waveform, as this could provide a way to constrain deviations from GR in the vicinity of BH.
These directions would pave the way for a deeper understanding of tidal responses in alternative theories of gravity and their potential observational signatures.

\section*{Acknowledgements}
V.Y.~is grateful for the hospitality of the external program APCTP-GW2025 held at Academia Sinica, Taipei, Taiwan, where part of this work was carried out.
This work was supported in part by World Premier International Research Center Initiative (WPI), MEXT, Japan.
The work of H.K.~was supported by JST (Japan Science and Technology Agency) SPRING, Grant No.\ JPMJSP2110.
The work of S.M.~was supported in part by JSPS (Japan Society for the Promotion of Science) KAKENHI Grant No.\ JP24K07017. 
The work of N.O.~was supported by JSPS KAKENHI Grant No.\ JP23K13111 and by the Hakubi project at Kyoto University.
The work of K.T.~was supported by JSPS KAKENHI Grant No.\ JP23K13101.
The work of V.Y.~is supported by grants for development of new faculty staff, Ratchadaphiseksomphot Fund, Chulalongkorn University and by the National Science, Research and Innovation Fund (NSRF) via the Program Management Unit for Human Resources \& Institutional Development, Research and Innovation Grant No.\ B39G680009.

\appendix

\section{Supplementary materials for MST method}\label{app:MST}

\subsection{\texorpdfstring{Low-frequency expansion of the MST coefficients~$a_n^\nu$}{Low-frequency expansion of the MST coefficients}}\label{app:order-an}
Here, we discuss the low-frequency expansion of the MST coefficients~$a_n^\nu$ in Eq.~(\ref{MST-hor-in}).
Using Eqs.~\eqref{alpha-beta-gamma} and \eqref{reccurence-for-RL}, we can evaluate the leading behaviors of $R_n(\nu)$ and $L_n(\nu)$ in the limit~$\varepsilon\ll 1$.
For positive $n$, we find that $R_n(\nu)\sim {\cal O}(\varepsilon)$, which gives $a_n^\nu\sim{\cal O}(\varepsilon^n)$.
Also, the behavior of $L_n(\nu)$ for negative integers $n$ are as follows:
\begin{equation}
    \begin{split}
    &L_{-\ell+1}(\nu)\sim {\cal O}(\varepsilon^3)\;,\quad
    L_{-\ell-1}(\nu)\sim {\cal O}(\varepsilon^0)\;,\quad
    L_{-2\ell-1}(\nu)\sim {\cal O}(\varepsilon^{-1})\;, \\
    &L_{n}(\nu)\sim {\cal O}(\varepsilon)\quad \text{for all other negative integers~$n$}\;.
    \end{split}
\end{equation}
Thus, the leading behaviors of the coefficients~$a_n^\nu$ in the low-frequency limit are given by
\begin{align}
    a_n^\nu\sim
    \left\{
    \begin{array}{rl}
         {\cal O}(\varepsilon^{|n|}) \;; & \text{for } -\ell+2\le n\;,\\
         {\cal O}(\varepsilon^{\ell+1}) \;;& \text{for } n=-\ell+1\;,\\
         {\cal O}(\varepsilon^{\ell+2}) \;;& \text{for } n=-\ell-1,\ -\ell\;,\\
         {\cal O}(\varepsilon^{|n|+1}) \;;& \text{for } -2\ell\le n\le -\ell-2\;,\\
         {\cal O}(\varepsilon^{|n|-1}) \;;& \text{for } n\le -2\ell-1 \;.
    \end{array}\right.\label{MST-order-coefficient}
\end{align}
Furthermore, in Table~\ref{tab:MST-coeff}, we present the coefficients~$a_n^\nu$ for $n = 0, \pm1, \pm2$ in the small-$\varepsilon$ limit.
\begin{table}[H]
    \centering
    \caption{List of the MST coefficients~$a_n^\nu$ for $n = 0, \pm1, \pm2$ at small $\varepsilon$~\cite{Mano:1996mf,Casals:2015nja}.}
    \label{tab:MST-coeff}
    \vskip2mm
    \begin{tabular}{|c||c|}\hline
         $n$& $a_n^\nu$  \\ \hline\hline
         $0$& 1\\ \hline 
         $1$& $\displaystyle -i\varepsilon\frac{(\ell+3)^2}{2 (\ell+1) (2\ell+1)}+\varepsilon ^2\frac{(\ell+3)^2}{2 (\ell+1)^2 (2 \ell+1)}+{\cal O}(\varepsilon^3)$ \rule[-5mm]{0mm}{12mm} \\ \hline
         $-1$ & $\displaystyle -i\varepsilon\frac{(\ell-2)^2 }{2 \ell (2 \ell+1)}- \varepsilon^2\frac{(\ell-2)^2}{2 \ell^2 (2 \ell+1)}+{\cal O}(\varepsilon^3)$ \rule[-5mm]{0mm}{12mm} \\ \hline
         $2$& $\displaystyle -\varepsilon ^2\frac{(\ell+3)^2 (\ell+4)^2}{4 (\ell+1)(2 \ell+1)(2\ell+3)^2}+{\cal O}(\varepsilon^3)$ \rule[-5mm]{0mm}{12mm} \\ \hline
         $-2$ & $\displaystyle -\varepsilon ^2\frac{(\ell-3)^2 (\ell-2)^2 }{4 \ell (2 \ell-1)^2 (2 \ell+1)}+{\cal O} (\varepsilon^3)$  \rule[-5mm]{0mm}{12mm} \\ \hline
    \end{tabular}
\end{table}

\subsection{Complex reflection rate}\label{app:reflection-MST}
Here, we provide additional calculations for the far-region behavior of the MST functions and derive the complex reflection rate.
We can obtain the asymptotic behavior of the horizon-ingoing solution around infinity from Eq.~\eqref{hor-in-coulomb-wave} as follows:
\begin{equation}
    \psi_\nu^{- {\rm in}}(x)
    \sim A_\nu^{\rm out}e^{i\varepsilon x}(\varepsilon x)^{i\varepsilon}+A_\nu^{\rm in}e^{-i\varepsilon x}(\varepsilon x)^{-i\varepsilon}\;,
\end{equation}
with
\begin{align}
\begin{split}
    A^{\rm out}_\nu&=e^{-\frac{\pi}{2}\varepsilon}2^{-1+i\varepsilon}[K_\nu(-i)^{\nu+1}+K_{-\nu-1}i^\nu]\left(\sum_{n=-\infty}^\infty  \frac{\Gamma(n+\nu-1-i\varepsilon)\Gamma(n+\nu+1-i\varepsilon)}{\Gamma(n+\nu+1+i\varepsilon)\Gamma(n+\nu+3+i\varepsilon)}a_n^\nu\right)\;,\\
    A^{\rm in}_\nu&=e^{\frac{\pi}{2}\varepsilon}2^{-1-i\varepsilon}\left[K_\nu i^{\nu+1}+K_{-\nu-1}(-i)^\nu\frac{\sin[\pi(\nu+i\varepsilon)]}{\sin[\pi(\nu-i\varepsilon)]}\right] \left(\sum_{n=-\infty}^\infty (-)^n \frac{\Gamma(n+\nu-1-i\varepsilon)}{\Gamma(n+\nu+3+i\varepsilon)}a_n^\nu\right)\;.
\end{split}
\end{align}
The complex reflection rate is defined as
\begin{equation}
    \mathcal{R}_\nu\equiv\frac{A^{\rm out}_\nu}{A^{\rm in}_\nu}=e^{-\pi\varepsilon}2^{2i\varepsilon}(-)^{\nu+1}\mathcal{C}_\nu \frac{\sum_{n=-\infty}^\infty  \frac{\Gamma(n+\nu-1-i\varepsilon)\Gamma(n+\nu+1-i\varepsilon)}{\Gamma(n+\nu+1+i\varepsilon)\Gamma(n+\nu+3+i\varepsilon)}a_n^\nu}{\sum_{n=-\infty}^\infty (-)^n \frac{\Gamma(n+\nu-1-i\varepsilon)}{\Gamma(n+\nu+3+i\varepsilon)}a_n^\nu}\;, \label{reflection-factor}
\end{equation}
where
\begin{equation}
\mathcal{C}_\nu = \frac{1+ie^{i\pi\nu}\frac{K_{-\nu-1}}{K_\nu}}{1-ie^{-i\pi\nu}\frac{\sin [\pi(\nu+i\varepsilon)]}{\sin [\pi(\nu-i\varepsilon)]}\frac{K_{-\nu-1}}{K_\nu}}\;.
\end{equation}
In the $\varepsilon\to 0$~limit, the complex reflection ratio behaves as $\mathcal{R}_\nu= A^{\rm out}_\nu/A^{\rm in}_\nu \sim (-)^{\nu+1}e^{-\pi\varepsilon}2^{2i\varepsilon}\Gamma(\nu+1+i\varepsilon)/\Gamma(\nu+1-i\varepsilon)$. 
It is crucial to mention that, in order to relate this reflection factor to the tidal response function, one needs to isolate the contributions of tidal effects to the scattering process~\cite{Saketh:2024juq,DeLuca:2024ufn}.
To do so, one possible procedure is to make use of the near-far factorization, as presented in~\cite{Bautista:2023sdf,Saketh:2024juq}. 
The factor~$\mathcal{C}_\nu$ represents the contribution from the boundary condition at the horizon, which depends on the ratio~$K_{-\nu-1}/K_\nu \sim{\cal O}(\varepsilon^{2\ell})$.
Essentially, the near-zone phase shift is the scattering against the star surface which contains the tidal response, whereas the other part corresponds to the wave scattering against the curved background metric.
Another approach we used in our parametrized formalism is to read off the tidal response function from the reflection coefficient, e.g.,~\cite{Bautista:2023sdf,Saketh:2024juq}.
Note, however, that in the case of parametrized formalism, both the far-zone solution and the connection formula need to be modified.

\subsection{\texorpdfstring{MST solutions for $\ell=2$}{MST solutions for l=2}}\label{app:MST-specific}
Here, for demonstration purposes, we derive the low-frequency expansion of the MST solutions for $\ell=2$.
In this case, for $x\gg 1$ and $\varepsilon x\ll 1$, the solutions~$\psi_\nu^{(0)}(x)$ and $\psi_{-\nu-1}^{(0)}(x)$ in Eqs.~(\ref{growing}) and (\ref{decaying}) take the form
\begin{equation}
\begin{split}
   \psi_\nu^{(0)}(x)
    &= i\varepsilon^{-1}\left[\frac{x^3}{24}+\frac{6+5/x}{1440\ell_2 x^2}+{\cal O}\left(x^{-4}\right)\right]\left(1+i\varepsilon\frac{6\ell_2-5}{6}\right)\;, \\
    \psi_{-\nu-1}^{(0)}(x)
    &= i\varepsilon^{-1}\left[-\frac{6+5/x}{1440\ell_2 x^2}+{\cal O}\left(x^{-4}\right)\right]\left(1-i\varepsilon\frac{6\ell_2+5}{6}\right)\;,
\end{split}
\end{equation}
where we have kept terms up to the first order in $\varepsilon$.
In what follows, for concreteness, let us restrict ourselves to the Hayward solution in the EFT of scalar-tensor gravity studied in Section~\ref{sec:EFT_gen_RW}.
In this case, the source term in Eq.~\eqref{perturbed-eq} can be written as
\begin{align}
    \delta V_2^-\psi_\nu^{-{\rm in}}(x)
   &= \eta \varepsilon^{-1}\bigg\{\frac{i}{12}+\frac{i}{16 x}+\frac{19 i}{24 x^2}-\frac{7 i}{16 x^3} + \varepsilon \bigg[\frac{1}{144} (10-12 \ell_2)+\frac{10-12 \ell_2}{192 x}  \nonumber \\
   &\hspace{4mm} -\frac{19 (12 \ell_2-10)}{288 x^2}+\frac{7 (12 \ell_2-10)}{192 x^3}+{\cal O}\left(x^{-4}\right)\bigg] \bigg\}\;, \label{Haywa-d-l=2-source}
\end{align}
at the leading order in $\varepsilon$.
Notice that the leading term in $\varepsilon$ is the same as that in Eq.~(3.68) of \cite{Barura:2024uog} up to a constant factor.
Therefore, the first-order solution~$\delta\psi_\nu^{-{\rm in}}(x)$ of Eq.~\eqref{eq:GRW-inhomogeneous-sol} in the low-frequency limit is
\begin{align}
    \ell_2\varepsilon^2 W_f\delta\psi_\nu^{-{\rm in}}(x) &= \eta\bigg\{ \frac{i x^2}{55296}+\frac{i x}{55296}+\frac{7 i}{55296}  +\varepsilon\bigg[\frac{(5-2 \ell_2) x^2}{110592}+\frac{(5-2 \ell_2) x}{110592} -\frac{7 (2 \ell_2-5)}{110592} \nonumber \\
   &\hspace{4mm} -\frac{1}{15360 x^2}-\frac{43}{829440 x^3}+{\cal O}(x^{-4})\bigg]+{\cal O}(\varepsilon^2) \bigg\}\;.
\end{align}
Note that this does not immediately correspond to the tidal response function. 
As discussed before, it is necessary to extract the part corresponding to the tidal response function from the constant part of each integral in Eq.~(\ref{delta-C_(1)}),
\begin{align}
\begin{split}
    W_f\delta C_+^-&=\frac{\eta}{\ell_2\varepsilon^2}\left[-\frac{1}{256}-\frac{i (540 \ell_2-101) \varepsilon }{34560}+{\cal O}(\varepsilon^2)\right]\;,\\
    W_f\delta C_-^-&= \frac{\eta}{\ell_2\varepsilon^2}\left[\frac{1}{256}-\frac{101 i\varepsilon}{34560}+{\cal O}(\varepsilon^2)\right]\;.
\end{split}
\end{align}
Therefore, taking into account the Wronskian factor and the coefficients of the MST functions in the definition (\ref{bare-psi}), the correction to the tidal response function at first-order in $\eta$ is 
\begin{align}
    \delta {\cal F}_\nu^{-,\,{\rm Bare}}(\varepsilon) &=\left(\delta C_+^-+\delta C_-^-\right)\times\frac{\left(\text{the coefficient of $x^{-\nu}$ in $\psi_{-\nu-1}^{(0)}(x)$}\right)}{\left(\text{the coefficient of $x^{\nu+1}$ in $\psi_{\nu}^{(0)}(x)$}\right)} \nonumber \\
    &=-\frac{24^2}{5}\left[1+i\varepsilon\frac{6\ell_2-5}{6}+{\cal O}(\varepsilon^2)\right]^{-2}\bigg[-i\varepsilon\frac{\eta}{64}+{\cal O}(\varepsilon^2)\bigg] \nonumber \\
    &=\frac{9}{5}\eta i\varepsilon+{\cal O}(\varepsilon^2)\;.
\end{align}
This gives the same result as that obtained in Eq.~(\ref{eq:TDN_ell=2_Hayward}) using the parametrized formalism.

\section{Logarithmic running of tidal Love numbers}\label{app:subtraction}
In this appendix, we discuss the origin of the logarithmic running of tidal Love numbers. 
Although in the main text we have focused on the TDNs, it is worth pointing the situation when one encounters the logarithmic running of TLNs.

Let us recall that in this paper we use the analytic continuation of $\ell$ to non-integer values ($\ell \to \nu =  \ell + \Delta \ell$) in order to distinguish between the source and response terms.
By doing so, the solution may develop a pole in the limit~$\Delta\ell\to 0$.
For the MST solution discussed in Sections~\ref{sec:review-tidal-response}, such a singularity appears in $\psi_{\nu+1}^{(0)}(x)$ as $C_\ell^{\rm div}(\varepsilon) x^{-\ell+\Delta\ell}/(2\Delta\ell)$ and in $\psi_{-\nu}^{(0)}(x)$ as $-C_\ell^{\rm div}(\varepsilon) x^{-\ell-\Delta\ell} /(2\Delta\ell)$, where the coefficient $C_\ell^{\rm div}$ starts at the second order in $\varepsilon$, i.e., $C_\ell^{\rm div}(\varepsilon)\sim{\cal O}(\varepsilon^2)$.
If we let $\Delta\ell = \ell_2\varepsilon^2+{\cal O}(\varepsilon^4)$, this would lead to a pole in the limit~$\ell_2 \to 0$.\footnote{One way to interpret this pole is through the worldline EFT framework, where it arises from divergences in loop diagrams and their corresponding counterterms when using dimensional regularization.}
Notice that the factor~$\ell_2$ does not appear in the denominator of the total horizon-ingoing solution.
In the case of vacuum 4d GR, the bare static TLNs are identically zero, so that the counterterm vanishes, i.e., $C_\ell^{\rm div}\to 0$ as $\varepsilon\to 0$.
This is indeed a consequence of the protection by hidden {\it ladder} symmetry~(see, e.g.,~\cite{Hui:2021vcv,BenAchour:2022uqo,Hui:2022vbh,Charalambous:2022rre,Katagiri:2022vyz,Berens:2022ebl}).

In the presence of deviations from vacuum GR, $C_\ell^{\rm div}(\eta, \varepsilon)$ is non-vanishing even in the static limit and, at the leading order, scales as $C_\ell^{\rm div}(\eta, \varepsilon)\sim{\cal O}(\eta, \varepsilon^2)$, where $\eta$ is a modified-gravity parameter.
Also, the fact that our system obeys the time-reversal symmetry implies that the coefficient~$C_\ell^{\rm div}(\eta, \varepsilon)$ is an even function of $\varepsilon$.
To explicitly see the logarithmic running, we consider $x^{-\ell}$ terms in $\psi_\nu^{-{\rm in}}(x)$,
\begin{align}\label{eq:limit_log}
   \frac{C_\ell^{\rm div}(\eta, \varepsilon)}{2\Delta\ell}x^{-\ell+\Delta\ell}-\frac{C_\ell^{\rm div}(\eta, \varepsilon)}{2\Delta\ell}x^{-\ell-\Delta\ell} = x^{-\ell}\left[C_\ell^{\rm div}(\eta, \varepsilon)\log x+{\cal O}(\Delta\ell)\right]\;,
\end{align}
at the leading order in $\Delta\ell$.
Note that the divergence at $\Delta\ell$ has been canceled.
In Eq.~(\ref{eq:limit_log}) terms of order~$\mathcal{O}(\Delta \ell)$ are finite as $\Delta \ell \to 0$, while the ${\cal O}((\Delta\ell)^0)$ coefficient contains the logarithmic dependence in $x$ which gives rise to the logarithmic running of the TLNs.\footnote{See \cite{Charalambous:2021mea} for a similar discussion on the dynamical TLNs of Kerr BH. Note that in the Kerr BH case, the leading real part of the tidal response starts at the first order in $\varepsilon$, which also has the logarithmic running part.}
We point out that, in the case of GR where $\eta = 0$, the logarithmic running appears at $(\varepsilon)^{2n}$, with $n$ being positive integers, i.e., no logarithmic running appears in the static TLNs.
On the other hand, in the case of modified gravity where $\eta\neq 0$, both the static and the dynamical TLNs generally contain the logarithmic running if $e^{\pm,\,j}_{(0)}$ diverge as $1/(2\Delta\ell)$.
Note that the logarithmic term, which is extracted from the coefficients of $1/(2\Delta\ell)$ (i.e., the singular part of $e_{(0)}^{-,\,j}$), is the same as that given in Appendix~B of \cite{Barura:2024uog}.

\section{EFT of black hole perturbations with timelike scalar profile}\label{app:EFT-BHPT}
In this appendix, we provide a brief review of the EFT of black hole perturbations with timelike scalar profile developed in Refs.~\cite{Mukohyama:2022enj,Mukohyama:2022skk}.
The key idea behind the formulation of the EFT is that, when a time-dependent background of the scalar field~$\Phi$ is chosen to coincide with the time coordinate~$\tau$ (i.e., in the unitary gauge), the 4d diffeomorphism invariance is spontaneously broken to the 3d diffeomorphism invariance.
This means that the EFT action contains any scalar functions made of 4d and 3d diffeomorphism covariant quantities, e.g., the 4d metric~$g_{\mu\nu}$, its inverse~$g^{\mu\nu}$, the 4d Riemann tensor~$R^{\mu\nu}{}_{\rho\sigma}$, the extrinsic curvature~$K^\mu_\nu$ of a constant-$\tau$ hypersurface, the time~$\tau$ itself and its derivative~$\partial_{\mu}\tau=\delta^{\tau}_{\mu}$. 
To apply the EFT framework to BH perturbations, we expand the action around a BH background which is spatially inhomogeneous, implying that the expansion coefficients (i.e., EFT parameters) depend on the spatial coordinates in a non-trivial manner.
Nevertheless, the 3d diffeomorphism invariance of the EFT can be preserved by imposing a set of consistency relations to the EFT parameters at each order in perturbations~\cite{Mukohyama:2022enj}.

In what follows, we consider odd-parity perturbations about a non-rotating BH background based on the EFT, following Refs.~\cite{Mukohyama:2022skk,Mukohyama:2023xyf}.\footnote{An EFT of black hole perturbations can also be constructed for vector-tensor gravity~\cite{Aoki:2023bmz}, which has recently been applied to the study of quasinormal modes~\cite{Tomizuka:2025dpy}.}
In particular, we present the master equation of the odd-parity perturbations (i.e., generalized Regge-Wheeler equation) and discuss its application to the tidal response of stealth Schwarzschild solutions.
See Refs.~\cite{Mukohyama:2023xyf,Konoplya:2023ppx,Barura:2024uog,Oshita:2024fzf} for applications of the generalized Regge-Wheeler equation to quasinormal mode frequencies, static TLNs, and greybody factors.
It is worth noting that the dynamics of spherical BH perturbations in the even-parity sector, including the so-called scordatura effect~\cite{Motohashi:2019ymr}, was recently studied in \cite{Mukohyama:2025owu}.

\subsection{Generalized Regge-Wheeler equation}\label{app:gRW}
Let us write the static and spherically symmetric background metric as follows:
    \begin{align}
    \bar{g}_{\mu\nu}\de x^\mu \de x^\nu 
    = -A(r)\de t^2 + \frac{\de r^2}{B(r)} + r^2(\de\theta^2 + \sin^2\theta\,\de\varphi^2)\;, \quad A(r)=B(r)\;, \label{SSSmetric}
    \end{align}
where we have assumed $g_{rr}=-g_{tt}^{-1}$ for simplicity.
The EFT action relevant to linear odd-parity perturbations in the unitary gauge is given by~\cite{Mukohyama:2022skk}\footnote{Notice that we disregard contributions from other operators like $M_6(r)\bar{K}^\mu_\nu\delta K^\nu_\rho\delta K^\rho_\mu$, which was previously discussed in \cite{Mukohyama:2024pqe}. Also, we do not consider parity-violating operators in the EFT.}
\begin{align}
	S_{\rm odd} = \int \de^4x \sqrt{-g} \bigg[\frac{M_\star^2}{2}R - \Lambda(r) - c(r)g^{\tau\tau} -\tilde{\beta}(r) K - \alpha(r)\bar{K}^{\mu}_\nu K^\nu_{\mu} + \frac{1}{2} M_3^2(r) \delta K^\mu_\nu \delta K^\nu_\mu \bigg] \;, \label{eq:EFT_action_odd}
\end{align}
where $M_\star^2$ is a constant, $R$ is the 4d Ricci scalar, a bar denotes the background value, and we have defined $K\equiv K^\mu_\mu$ and $\delta K^\mu_\nu\equiv K^\mu_\nu-\bar{K}^\mu_\nu$.
The $r$-dependent coefficient functions correspond to the EFT parameters, among which $\Lambda$, $c$, $\tilde{\beta}$, and $\alpha$ are subject to tadpole cancellation conditions to ensure consistency with the background metric~\eqref{SSSmetric} (see Refs.~\cite{Mukohyama:2022enj,Mukohyama:2022skk} for details).
Also, we impose the condition~$\alpha+M_3^2=0$, as otherwise a slowly rotating BH solution does not exist or the radial sound speed diverges at spatial infinity~\cite{Mukohyama:2022skk}.
Note in passing that this condition is realized in many covariant higher-order scalar-tensor theories~\cite{Mukohyama:2022enj,Mukohyama:2022skk,Kobayashi:2025evr}.
In this setup, it was shown in Refs.~\cite{Mukohyama:2022skk,Mukohyama:2023xyf} that the master equation for the odd-parity perturbations (i.e., the generalized Regge-Wheeler equation) can be written in the form,
    \begin{align}
    \left[\left(F\frac{\partial}{\partial r}\right)^2-\frac{\partial^2}{\partial\tilde{t}^2}-F\tilde{V}_\ell^-\right]\tilde{\psi}_\ell^-=0\;, \label{gRW}
    \end{align}
where we have defined the following quantities:
    \begin{align}
    \begin{split}
    &F(r)\equiv \frac{A+\alpha_T}{\sqrt{1+\alpha_T}}\;, \quad
    \alpha_T(r)\equiv \frac{\alpha(r)}{M_\star^2-\alpha(r)}\;, \quad
    \tilde{t}\equiv t+ \int \frac{\sqrt{1-A}}{A}\frac{\alpha_T}{A+\alpha_T} {\rm d}r\;, \\
    &\tilde{V}_\ell^-(r)
    =\sqrt{1+\alpha_T}\,F\left\{\frac{\ell(\ell+1)-2}{r^2}+\frac{r}{(1+\alpha_T)^{3/4}}\left[ F\left(\frac{(1+\alpha_T)^{1/4}}{r}\right)'\,\right]'\right\}\;,
    \end{split}\label{gRW_building_blocks}
    \end{align}
with a prime denoting the derivative with respect to $r$.
Here, in the present setup, the functional form of $\alpha(r)$ is determined from the tadpole cancellation conditions as~\cite{Mukohyama:2023xyf}
    \begin{align}
    \alpha=M_\star^2+\frac{3C}{r(2-2A+rA')}\;, \label{sol_alpha}
    \end{align}
where $C$ is an integration constant.
Note that the position of the odd-mode horizon is determined by $F(r)=0$.
It should also be noted that the function~$\alpha_T(r)$ corresponds to (local) deviation of the propagation speed of GWs from that of light.
In the frequency domain, Eq.~\eqref{gRW} takes precisely the form~\eqref{deformed-RWZ0}, as the frequency conjugate to $\tilde{t}$ coincides with that conjugate to $t$.

\subsection{Tidal response of stealth Schwarzschild background}\label{app:stealth}
In scalar-tensor theories (or other modified gravity theories), the solutions of the metric can in general differ from those of GR.
However, there is a special class of solutions where the metric has the same form as in GR and the effect of a non-trivial scalar profile is hidden, at least at the background level.
Such solutions are called stealth solutions~\cite{Mukohyama:2005rw,Motohashi:2018wdq,Takahashi:2020hso}, and perturbations around them have been studied within both specific model or EFT approach, e.g.,~\cite{Babichev:2018uiw,Takahashi:2019oxz,deRham:2019gha,Motohashi:2019ymr,Khoury:2020aya,Tomikawa:2021pca,Takahashi:2021bml}.\footnote{It is known that the perturbations about stealth solutions in Horndeski and DHOST theories are strongly coupled in the asymptotic Minkowski/de Sitter region due to the vanishing of sound speed~\cite{deRham:2019gha,Khoury:2020aya,Takahashi:2021bml}.
The scordatura mechanism resolves this issue by introducing a higher-derivative term in the dispersion relation~\cite{Motohashi:2019ymr}, which can be naturally included as higher-derivative operators in the context of the EFT.
The strongly coupled mode is relevant only to the even-parity sector, and therefore the generalized Regge-Wheeler equation can be safely used to study the dynamics of odd-parity perturbations.}
Hereafter, we show that the tidal response for odd-parity perturbations of the stealth Schwarzschild background deviates from the GR values only by a constant scaling factor.

For a stealth Schwarzschild solution, the background metric function is given by
\begin{align}
A(r) =1-\frac{r_s}{r}\;,
\end{align}
where $r_s$ is a constant.
Note that $r_s$ is in general different from the horizon radius~$r_g$ of the odd-parity perturbations.
In this case, the parameter~$\alpha_T$ in Eq.~\eqref{gRW_building_blocks} is a constant, and the generalized Regge-Wheeler equation~\eqref{gRW} in the frequency domain becomes
\begin{align}
\left[\left(F(r)\frac{\de}{\de r}\right)^2+\omega^2-F(r)\tilde{V}_\ell^-(r)\right]\tilde{\psi}_\ell^-(r;\omega)=0\;, \quad
\tilde{V}_\ell^-(r)=\sqrt{1+\alpha_T}\left(\frac{\ell(\ell+1)}{r^2}-\frac{3r_g}{r^3}\right)\;,
\label{genRW_stealth}
\end{align}
with $F(r)=\sqrt{1+\alpha_T}(1-r_g/r)$ and $r_g=r_s/(1+\alpha_T)$.
Interestingly, Eq.~\eqref{genRW_stealth} is equivalent to the standard Regge-Wheeler equation~\eqref{GR-RW} in GR up to a rescaling of $\omega$, i.e.,
    \begin{align}
    \tilde{\omega}\equiv\frac{\omega}{\sqrt{1+\alpha_T}}\;. \label{tilde_omega}
    \end{align}
Indeed, after applying the rescaling, the generalized Regge-Wheeler equation takes the form,
\begin{align}
\left[\left(f(r)\frac{\de}{\de r}\right)^2 +\tilde{\omega}^2-f(r)V_\ell^-(r)\right]\tilde{\psi}_\ell^-(r;\sqrt{1+\alpha_T}\,\tilde{\omega})=0\;, \quad
V_\ell^-(r)=\frac{\ell(\ell+1)}{r^2}-\frac{3r_g}{r^3}\;,
\end{align}
which is nothing but the Regge-Wheeler equation in GR.
This suggests that the linear dynamics of the odd-parity modes on a stealth Schwarzschild background is the same as that in GR up to a simple rescaling~\cite{Mukohyama:2023xyf,Nakashi:2023vul}.
This immediately implies that the static TLNs of a stealth Schwarzschild BH are vanishing as in the GR case~\cite{Barura:2024uog}.

Let us now study the tidal response for $\omega\ne 0$.
To reiterate, in terms of the rescaled frequency~$\tilde{\omega}$ in Eq.~\eqref{tilde_omega}, we have recast the master equation in the form of the Regge-Wheeler equation in GR.
Therefore, the (renormalized) tidal response function can be written in the same form as in GR, i.e.,
\begin{align}
    {\cal F}_\ell^{-,\,{\rm Ren}} = ir_g\tilde{\omega}{\cal N}_{\ell,{\rm GR}}+(r_g\tilde{\omega})^2{\cal K}^{(2)}_{\ell,{\rm GR}}+\cdots \;, \label{eq:response_stealth}
\end{align}
where ${\cal N}_{\ell,{\rm GR}}$ and ${\cal K}^{(2)}_{\ell,{\rm GR}}$ denote the GR values of the TDNs and the second-order (renormalized) dynamical TLNs, respectively, and the ellipsis refers to terms higher-order in frequency.
Notice that in Eq.~(\ref{eq:response_stealth}) we have set the static TLNs to zero.
Going back to the original $\omega$ via Eq.~\eqref{tilde_omega}, we have
\begin{align}
    {\cal F}_\ell^{-,\,{\rm Ren}} = ir_s\omega\frac{{\cal N}_{\ell,{\rm GR}}}{(1+\alpha_T)^{3/2}}+(r_s\omega)^2\frac{{\cal K}^{(2)}_{\ell,{\rm GR}}}{(1+\alpha_T)^3}+\cdots \;,
\end{align}
where we have used the Schwarzschild horizon radius~$r_s$ instead of the odd-mode horizon radius~$r_g$, as $r_s$ can be probed through the motion of matter (such as electromagnetic waves, stars and gasses) and is therefore more relevant to observations.
We then read off both the TDNs and the second-order dynamical TLNs as
\begin{align}
{\cal N}_\ell^-= \frac{{\cal N}_{\ell,{\rm GR}}}{(1+\alpha_T)^{3/2}}\;, \quad
{\cal K}_\ell^{(2)-} = \frac{{\cal K}^{(2)}_{\ell,{\rm GR}}}{(1+\alpha_T)^3}\;. 
\end{align}
Hence, when $\alpha_T < 0$ and $r_g > r_s$ (subluminal propagation), we have ${\cal N}_\ell^- > {\cal N}_{\ell,{\rm GR}}$ and $\big|{\cal K}_\ell^{(2)-}\big| > \big|{\cal K}^{(2)}_{\ell,{\rm GR}}\big|$.
On the other hand, when $\alpha_T > 0$ and $r_g < r_s$ (superluminal propagation), ${\cal N}_\ell^-$ and $\big|{\cal K}_\ell^{(2)-}\big|$ are smaller than their corresponding GR values.

\small
\bibliographystyle{utphys}
\bibliography{bib}

\end{document}